\RequirePackage{rotating}
\documentclass[iop]{emulateapj} 
\usepackage{xspace}
\usepackage{natbib}
\usepackage{hhline}
\usepackage{amsmath}
\usepackage{dcolumn}
\usepackage{verbatim}
\usepackage{calc}
\usepackage{color}
\usepackage{booktabs}
\usepackage{longtable}
\usepackage[bookmarks]{hyperref}
\usepackage[numbered]{bookmark}
\usepackage{rotating}
\usepackage{paralist}
\usepackage{tabto}
\usepackage[multidot]{grffile}

\newcolumntype{R}{@{\extracolsep{5pt}}r@{\extracolsep{3pt}}}%
\newcolumntype{L}{@{\extracolsep{5pt}}l@{\extracolsep{3pt}}}%
\newcolumntype{C}{@{\extracolsep{5pt}}c@{\extracolsep{3pt}}}%

\newcolumntype{.}{D{.}{.}{-1}}

\newcommand{\chandra}{{\it Chandra}\xspace}
\newcommand{\suzaku}{{\it Suzaku}\xspace}

\newcommand{\swift}{{\it Swift}\xspace}
\newcommand{\xmm}{{\it XMM-Newton}\xspace}
\newcommand{\integral}{{\it INTEGRAL}\xspace}
\newcommand{\fermi}{{\it Fermi}\xspace}

\newcommand{\nustar}{{\it NuSTAR}\xspace}

\newcommand{\sS}[1]{\mbox{$\rm{}^{#1}$}}
\newcommand{\Ss}[1]{\mbox{$\rm{}_{#1}$}}
\newcommand{\nep}[2]{\mbox{${#1}$$\times$${10}^{#2}$}}

\newcommand{\Ms}{\mbox{$M_{\odot}$}\xspace}

\newcommand{\nH}{\mbox{$N$\Ss{H}}\xspace}

\newcommand{\lnls}{{log$N$-log$S$}\xspace}
\newcommand{\Deg}{\mbox{$^\circ$}\xspace}
\newcommand{\x}{\mbox{$\times$}}

\newcommand{\lcgs}{\mbox{erg s\sS{-1}}\xspace}
\newcommand{\fcgs}{\mbox{erg  s\sS{-1} cm\sS{-2}}\xspace}
\newcommand{\pcgs}{\mbox{ph cm\sS{-2} s\sS{-1}}\xspace}

\begin{document}

\title{\nustar Hard X-ray Survey of the Galactic Center Region II: 
X-ray Point Sources}



\author{
JaeSub Hong\altaffilmark{1}, 
Kaya Mori\altaffilmark{2}, 			
Charles J.~Hailey\altaffilmark{2}, 		
Melania Nynka\altaffilmark{2}, 		
Shuo Zhang\altaffilmark{2}, 		
Eric Gotthelf\altaffilmark{2},		
Francesca M.~Fornasini\altaffilmark{3},		
Roman Krivonos\altaffilmark{5}, 		
Franz Bauer\altaffilmark{6,7,8},		
Kerstin Perez\altaffilmark{9}, 		
John A.~Tomsick\altaffilmark{4}, 		
Arash~Bodaghee\altaffilmark{10},		
Jeng-Lun Chiu\altaffilmark{4}, 		
Ma{\"i}ca Clavel\altaffilmark{4},		
Daniel Stern\altaffilmark{11},		
Jonathan E.~Grindlay\altaffilmark{1}, 		
David M.~Alexander\altaffilmark{12},	
Tsuguo Aramaki\altaffilmark{13},		
Frederick K.~Baganoff\altaffilmark{14}, 		
David~Barret\altaffilmark{15,16}, 	
Nicolas Barri\`ere\altaffilmark{4}, 	
Steven E.~Boggs\altaffilmark{4}, 
Alicia M.~Canipe\altaffilmark{2},		
Finn E.~Christensen\altaffilmark{15}, 
William~W.~Craig\altaffilmark{4,18}, 
Meera A.~Desai\altaffilmark{2},		
Karl Forster\altaffilmark{19}, 
Paolo Giommi\altaffilmark{20}, 
Brian W.~Grefenstette\altaffilmark{19},
Fiona A.~Harrison\altaffilmark{19}, 
Dooran Hong\altaffilmark{2},
Allan Hornstrup\altaffilmark{17}, 
Takao Kitaguchi\altaffilmark{21,22}, 
Jason E.~Koglin\altaffilmark{23},
Kristen K.~Madsen\altaffilmark{19}, 
Peter H.~Mao\altaffilmark{19}, 
Hiromasa Miyasaka\altaffilmark{19}, 
Matteo Perri\altaffilmark{20, 24},
Michael J.~Pivovaroff\altaffilmark{18}, 
Simonetta Puccetti\altaffilmark{20,24}, 
Vikram Rana\altaffilmark{19}, 
Niels J.~Westergaard\altaffilmark{17},
William W.~Zhang\altaffilmark{25} 
and Andreas~Zoglauer\altaffilmark{4} \\
}
\altaffiltext{1}{Harvard-Smithsonian Center for Astrophysics, 60 Garden St., Cambridge, MA 02138, USA; jaesub@head.cfa.harvard.edu} 
\altaffiltext{2}{Columbia Astrophysics Laboratory, Columbia University, New York, NY 10027, USA} 
\altaffiltext{3}{Astronomy Department, University of California, Berkeley CA 94720, USA}
\altaffiltext{4}{Space Sciences Laboratory, University of California, Berkeley, CA 94720, USA} 
\altaffiltext{5}{Space Research Institute, Russian Academy of Sciences, Profsoyuznaya 84/32, 117997 Moscow, Russia} 
\altaffiltext{6}{Instituto de Astrof\'isica, Facultad de F\'isica, Pontificia Universidad Catolica de Chile, 306, Santiago 22, Chile} 
\altaffiltext{7}{Millennium Institute of Astrophysics, Santiago, Chile} 
\altaffiltext{8}{Space Science Institute, 4750 Walnut Street, Suite 205, Boulder, Colorado 80301} 
\altaffiltext{9}{Haverford College, 370 Lancaster Avenue, KINSC L109, Haverford, PA 19041, USA} 
\altaffiltext{10}{Georgia College, 231 W. Hancock St., Milledgeville, GA 31061, USA} 
\altaffiltext{11}{Jet Propulsion Laboratory, California Institute of Technology, Pasadena, CA 91109, USA } 
\altaffiltext{12}{Department of Physics, Durham University, Durham DH1 3LE, UK} 
\altaffiltext{13}{Standford National Accelerator Laboratory, 2575 Sand Hill Road, Menlo Park, CA 94025, USA}
\altaffiltext{14}{Kavli Institute for Astrophysics and Space Research, Massachusets Institute of Technology, Cambridge, MA 02139, USA} 
\altaffiltext{15}{Universit\'e de Toulouse, UPS-OMP, IRAP, Toulouse, France} 
\altaffiltext{16}{CNRS, Institut de Recherche en Astrophysique et Plan\'etologie, 9Av. colonel Roche, BP 44346, F-31028 Toulouse Cedex 4, France} 
\altaffiltext{17}{DTU Space - National Space Institute, Technical University of Denmark, Elektrovej 327, 2800 Lyngby, Denmark} 
\altaffiltext{18}{Lawrence Livermore National Laboratory, Livermore, CA 94550, USA} 
\altaffiltext{19}{Cahill Center for Astronomy and Astrophysics, California Institute of Technology, Pasadena, CA 91125, USA} 
\altaffiltext{20}{ASI Science Data Center, Via del Politecnico snc I-00133, Roma, Italy} 
\altaffiltext{21}{Department of Physical Science, Hiroshima University, Higashi-Hiroshima, Hiroshima 739-8526, Japan} 
\altaffiltext{22}{Core of Research for the Energetic Universe, Hiroshima University, Higashi-Hiroshima, Hiroshima 739-8526, Japan} 
\altaffiltext{23}{Kavli Institute for Particle Astrophysics and Cosmology, SLAC National Accelerator Laboratory, Menlo Park, CA 94025, USA} 
\altaffiltext{24}{INAF - Astronomico di Roma, via di Frascati 33, I-00040 Monteporzio, Italy} 
\altaffiltext{25}{NASA Goddard Space Flight Center, Greenbelt, MD 20771, USA} 


\begin{abstract} 
We present the first survey results of hard X-ray point sources in
the Galactic Center (GC) region by \nustar. We have discovered 70 hard
(3--79 keV) X-ray point sources in a 0.6 deg\sS{2} region around Sgr~A*
with a total exposure of 1.7 Ms, and 7 sources in the Sgr~B2 field with 300 ks.
We identify clear \chandra counterparts for 58 \nustar sources and assign
candidate counterparts for the remaining 19. The \nustar survey reaches
X-ray luminosities of $\sim$4$\times$ and $\sim$8$\times$10\sS{32}
\lcgs at the GC (8 kpc) in
the 3--10 and 10--40 keV bands, respectively. The source list includes three
persistent luminous X-ray binaries and the likely run-away pulsar called
the Cannonball.  New source-detection significance maps 
reveal a cluster of hard ($>$10 keV) X-ray sources near the Sgr~A diffuse
complex with no clear soft X-ray counterparts.  The severe extinction
observed in the \chandra spectra indicates that all the \nustar sources are
in the central bulge or are of extragalactic origin. 
Spectral analysis of relatively bright 
\nustar sources suggests that magnetic cataclysmic variables constitute
a large fraction ($>$40--60\%).
Both spectral analysis and \lnls distributions of the \nustar sources
indicate that the X-ray spectra of the \nustar sources should have $kT$
$>$ 20 keV on average for a single temperature thermal plasma model or an
average photon index of $\Gamma$ $=$ 1.5 -- 2 for a power-law model.  
These findings suggest that the GC X-ray source population may contain
a larger fraction of X-ray binaries with high plasma temperatures than
the field population. 
\end{abstract}

\section{Introduction\label{s:instro}}

The high density stellar cluster around the super massive black hole 
at the center of the Milky Way is of great interest for galaxy formation
and evolution processes because of its close proximity enabling
studies of individual stars, and because of the
likely ubiquity of such systems in the Universe. 
Since their discovery the nature of the thousands of X-ray sources
around Sgr~A* has long been the subject of extensive investigations 
beginning with \citet{Muno03}. Direct
identification of the X-ray sources in the Galactic Center (GC) region
through followup
optical/infrared imaging and spectroscopy has been difficult because
of severe obscuration ($A_V$$>$ 25), faint counterparts at large distances
($\sim$8 kpc), and source crowding \citep[e.g.][]{Berg09}. At a minimum, therefore,
a huge investment of time on large telescopes with adaptive
optics is required to overcome some of these challenges.  Nonetheless, high mass
X-ray binaries (HMXBs) were ruled out for
a majority early on \citep{Laycock05}: the lack of bright ($K$$<$15)
near-infrared (nIR) counterparts, which are expected from Be stars,
the most common companions in HMXBs, indicates that less than 5\% of the
X-ray sources in the GC region are HMXBs \citep{Mauerhan10}. 

A dominant source type of the X-ray sources in the GC region is currently believed
to be magnetic cataclysmic variables (MCVs), in particular,
intermediate polars (IPs), which fit the observed luminosity range
($L_X$$\sim$10\sS{31-33} \lcgs in 2--10 keV) and the unusually hard
X-ray spectra with equivalent power-law photon 
indices\footnote{In this paper, $\Gamma$ and $\Gamma$\Ss{S} are used to
describe the photon index for a power-law model in the broadband
($\sim$ 3--40 keV) and soft ($<$ 10 keV) bands, respectively.}
of $\Gamma$\Ss{S} $\sim$ 1 in the
2--10 keV band
\citep{Muno03, Hong09}. 
Active binaries (ABs) have been suggested to make a significant
contribution \citep{Revnivtsev09, Revnivtsev11}, although this has
been disputed \citep{Hong12}. 
\citet{Perez15} recently discovered apparent diffuse
hard X-ray emission (20--40 keV) in the central 2\arcmin\ region around Sgr~A* 
using \nustar observations.
Although the exact origin of the hard X-ray emission is not clear,
a leading scenario is that it is from an unresolved population of
1,000 -- 10,000 IPs with relatively high mass ($>$0.8 \Ms) white dwarfs (WDs), 
which explains a hard thermal component ($kT$$>$35 keV) observed in
the spectra \citep[see also][]{Hailey16}.  
MCVs are indeed likely to be a major
component of the X-ray source population at the GC,
given their higher abundance relative to black hole (BH) or neutron
star (NS) X-ray binaries (XBs), but a large population of quiescent,
exotic BH or NS systems cannot be ruled out yet.

To shed light on the nature of the X-ray source population in the
GC, we have surveyed the GC region around Sgr~A* using \nustar
\citep{Harrison13}.  \citet{Mori15} present the results from the \nustar
observations of the diffuse hard X-ray emission in the central 0.2 deg\sS{2}
region around Sgr~A*.  
In this paper, we report the first survey results and catalog of hard
X-ray {\it point} sources in a 0.6 deg\sS{2} region around Sgr~A* and a 0.06
deg\sS{2} region around the Sgr~B2 cloud.  With an angular resolution of
18\arcsec\ in Full-Width Half-Maximum (FWHM), \nustar is the only hard
X-ray telescope capable of resolving X-ray point sources in the crowded
GC region.  For comparison, previous hard X-ray emission above 10 keV
in the same region has been resolved into only three separate sources by \integral
\citep{Belanger06}.

\nustar studies of several prominent sources in the GC region
are found in the literature or underway: e.g.~CXO J174545.5--285828, a
likely run-away pulsar, aka the Cannonball by \citet{Nynka13}; 
1E 1743.1--2843, a possible NS low mass X-ray binary (LMXB) by
\citet{Lotti15}.  The \nustar observations 
of bright X-ray flares and bursts in the GC region are presented elsewhere: 
see \citet{Mori13} for the \nustar detection of X-ray pulsations from
SGR J1745-29, a transient magnetar \citep[see also][]{Kaspi14};
\citet{Barriere14} for the Sgr~A* flares; \citet{Barriere15} for 
Type I X-ray bursts from GRS1741.9--2853, a NS-LMXB; \citet{Younes15}
for an outburst from GRO J1744--28, also known as the Bursting Pulsar.

\nustar studies of a few bright diffuse sources in the GC region
are also found in the 
literature: \citet{Zhang14} and \cite{Nynka15}
report detailed \nustar hard X-ray studies of the X-ray filaments
Sgr~A-E knot and G359.97--0.038, respectively;
\citet{Krivonos14} and \citet{Zhang15} present the detailed \nustar
analysis results of the diffuse hard X-ray emission from the Arches cluster
and the Sgr~B2 cloud, respectively.
Complementing the survey of the GC region, a similar \nustar survey 
was conducted on the Norma sprial arm region in parallel. Some initial results
are reported in \citet{Bodaghee14}, while more complete coverage with
an in-depth analysis is underway \citep{Fornasini16}.

In \S2 we review the survey strategies and the \nustar observations of
the GC region and outline the data processing and mosaicking procedures. 
In \S3, we introduce a new source search
technique: after motivating the need for a new technique
(\S\ref{s:motivation}), we demonstrate how to build detection significance
maps called `trial maps' based on Poisson statistics-driven random chance probabilities
(\S\ref{s:trialmap}); we set detection thresholds 
by cross-correlating the trial maps with a \chandra source catalog (\S\ref{s:thresh})
and then present the \nustar source catalog (\S\ref{s:catalog}).  
In \S\ref{s:ap}, we present the aperture photometry of the
\nustar sources: we describe the aperture selection scheme
(\S\ref{s:aperture}) and summarize the photometry results
(\S\ref{s:photometry}), followed by the detailed description of
spectral classification (\S\ref{s:class}), 
flux and luminosity calculation (\S\ref{s:flux}),
spectral model fitting (\S\ref{s:fit}) 
and X-ray variability analysis (\S\ref{s:var}).
In \S5, we review the properties of several bright \nustar sources. 
In \S6, we explore unusually hard X-ray sources 
(\S\ref{s:hardsources}) and assess the survey
sensitivity (\S\ref{s:sensitivity}). We address
the lack of the foreground \nustar sources in our survey (\S\ref{s:missing})
and derive the \lnls distribution of the \nustar sources (\S\ref{s:lnls}).
Finally regarding the nature of the \nustar sources, we explore two
scenarios in addition to NS or BH XBs: MCVs (\S\ref{s:mcv}) and
rotationally powered pulsars (\S\ref{s:msp}).

\section{Observations and Data Processing}

\subsection{Observations and Data Screening}

Observations of the GC region with \nustar began in July, 2012, shortly
after launch.  
The original survey strategy for the GC region was to match
the central 2\Deg \x\ 0.7\Deg\ region covered by the \chandra X-ray
Observatory \citep[][hereafter M09]{Wang02, Muno09}.  
The field of views (FoVs) of neighboring \nustar observations in the
survey were designed to overlap with each
other by $\sim$40\%.  Multiple observations of the same region
with relatively large FoV offsets tend to average out the vignetting effects
of each observation, enabling a more uniform coverage of the region.
Multiple observations are also suitable for monitoring long term X-ray
variability of sources in the region.
Even when observing a single target, the \nustar
observation is often broken up into two or more segments with relatively
large pointing offsets to allow an efficient subtraction of a
detector coordinate-dependent background component \citep[e.g.][]{Mori13}. 

Our analysis includes three observations of the central Sgr~A* field.
The observations dedicated to the survey started with a coverage of a
0.2 deg\sS{2} region around Sgr~A*, which is called the mini-survey and served
as a pilot study. The rest of the 2\Deg \x\ 0.7\Deg\ region was divided
into four blocks.  It quickly became clear that more than half of the
survey region is dominated by stray light (SL)
or ghost ray (GR) background from nearby bright sources.  SL photons
register in the detectors without reflection from the \nustar optics,
arriving from large off-axis angles ($\sim$ 1 -- 5\Deg) through the
open gap between the optics structure and the detector collimators.
SL from a bright source forms a circular shaped region of high background
in the detectors.  
GRs arise from photons which only reflect once off the optics
("single-bounce"). A bright GR source generates a set of
radial streaks of high background in the detectors.
These background components are dominant at low
energies, below 10--40 keV, while the internal background becomes
dominant above 40 keV; see \citet{Wik14} and \citet{Mori15} for more details.


\begin{figure} \begin{center}
\includegraphics*[width=0.45\textwidth,clip=true]
{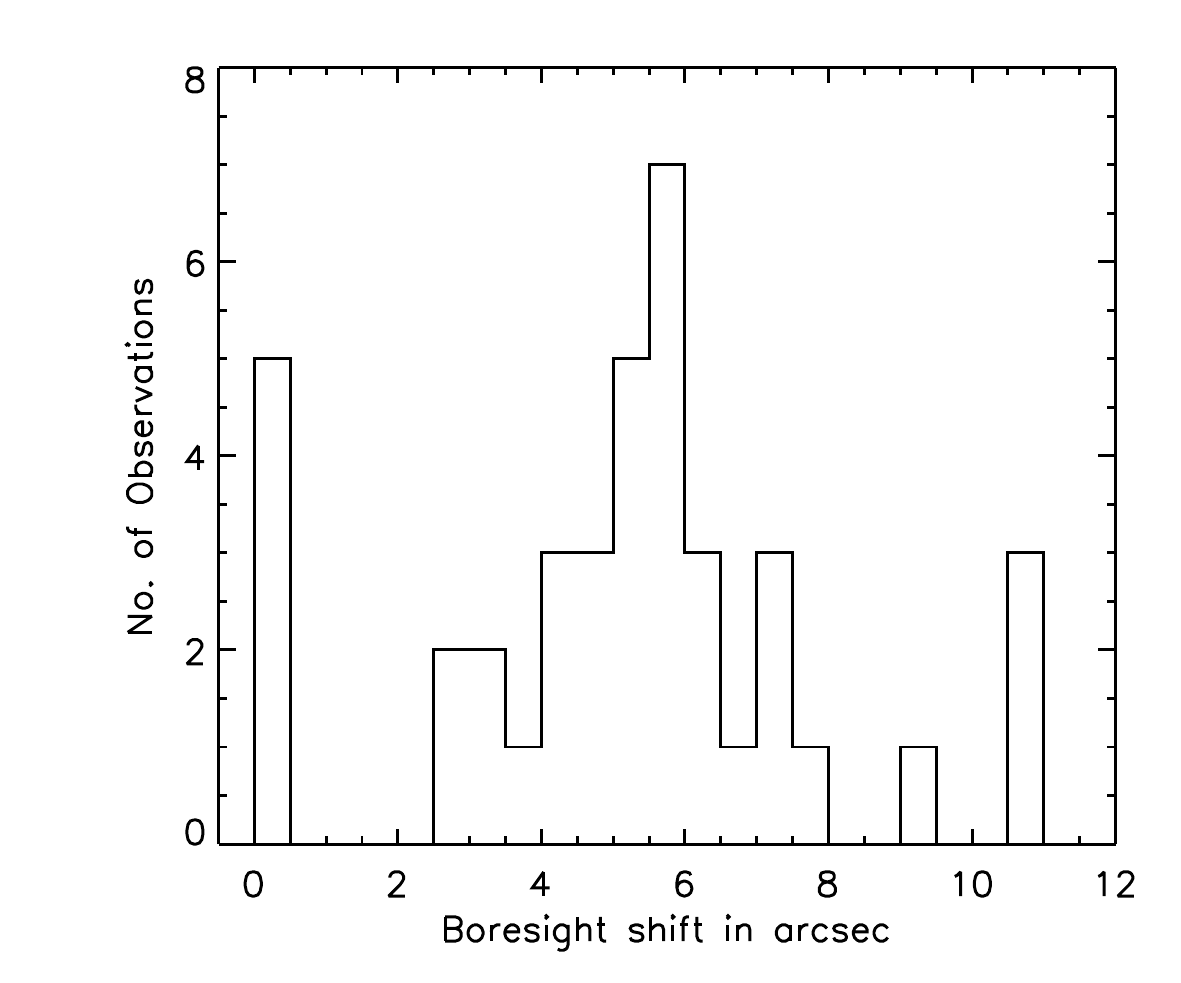}
\caption{Distribution of the boresight shifts applied for astrometric
correction before mosaicking the individual observations. 
For five observations with no clear bright sources to measure
boresight shifts, no astrometric correction is applied.  See
\S\ref{s:data} and Table~\ref{t:obs}.
}
\label{f:bs}
\end{center}
\end{figure}

We terminated our survey after coverage of the first two blocks (A and B)
because of the severe SL and GR backgrounds expected in the rest of the
region. Blocks A and B covered the Galactic northern and western
sides of the mini-survey region, respectively.  As a result,
a continuous 1.2\Deg \x\ 0.5\Deg region around Sgr~A* is covered by the survey.
In addition, our analysis includes two targeted deep observations of
the Sgr~B2 cloud. These observations were conducted to measure the
hard X-ray emission spectrum from the Sgr~B2 cloud and its temporal
change in morphology, and they also suffer from severe SL backgrounds.
Analysis of the cloud is found in \citet{Zhang15}, and here we report on the
X-ray point sources found in the field and their properties.


\begin{figure*} \begin{center}
\includegraphics*[width=0.97\textwidth,clip=true] {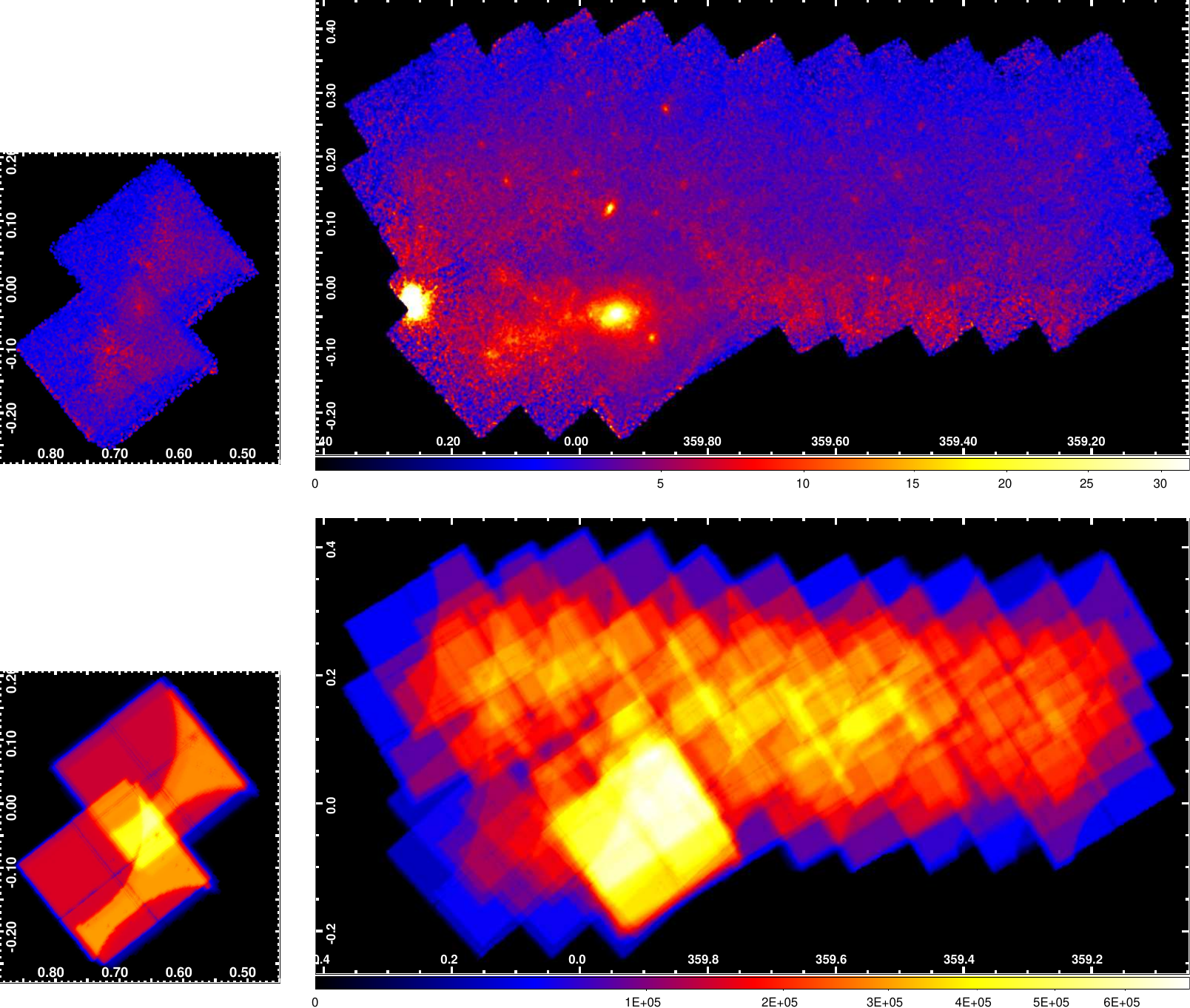}
\caption{(Top) Exposure-corrected smoothed \nustar images of
the main GC region (right) and the Sgr B2 cloud field (left) 
in the 10--40 keV band.
Smoothing is done via a Gaussian kernel of a four pixel radius in
SAOImage DS9.
(Bottom) Vignetting-free exposure mosaic of the same regions (not smoothed).
The $x$ and $y$ axes are Galactic longitude and latitude, respectively. The
color scale of the smoothed image is in counts s\sS{-1} pix\sS{-1},
and the exposure map in seconds. 
}
\label{f:raw}
\end{center}
\end{figure*}

Table~\ref{t:obs} summarizes the \nustar coverage of the GC region analyzed
in this paper.  
The depths of the various observations are as follows: Sgr~A* for 50--160
ks, Sgr~B2 for 160 ks,  mini survey region for 25 ks, and
blocks A and B for 40 ks.  We excluded X-ray events in the
self-evident SL patterns from our analysis.  Table~\ref{t:obs} lists the
focal plane modules (FPMs) whose SL patterns, if any, were removed. 
For instance, many observations in block B show bright SL backgrounds
in both modules.  We also excluded the data when SGR~J1745--29 was in outburst
(i.e.~only three observations of the Sgr~A* field were included) and the
burst data from GRS~1741.9--2853 (352 s around the peak of the burst 
from Obs.~ID 40010001002), as well
as strong and mild flares from Sgr~A* (40 ks from Obs.~ID 30001002001)
in order to improve detection sensitivity of nearby faint point sources
(see Table~\ref{t:obs}).

\subsection{Data Processing and Mosaicking} \label{s:data}

We processed the raw data of each observation to produce event files
and exposure maps for both of the \nustar modules (FPMA \& B) using the standard
\nustar pipeline v1.3.1 provided under HEASOFT v6.15.1.  The exposure
maps used in our analysis, except for non-parametric flux estimations
(\S\S\ref{s:flux} and \ref{s:lnls}), were generated without vignetting effects.
For both source detection (\S\ref{s:det}) and aperture photometry (\S\ref{s:ap}), we
used apertures symmetric with respect to the source position 
(albeit of different sizes), and thus the vignetting effects
are roughly averaged out to first order. 

Initially we attempted to localize the positions of a few brightest
sources in each observation for astrometric correction of the event
files and the exposure maps. This approach did not produce reliable
boresight shifts due to relatively high background and lack of bright
point sources in individual observations.  Instead we use the detection
significance map called 'trial map'
(see \S\ref{s:trialmap}), which are generated from the merged image and
exposure map (see below) without boresight correction.  For the main GC
region, we identified 14 bright sources in the trial maps of the 3--10
and 10--40 keV bands with clear \chandra counterparts, and localized
their positions using the IDL 2-D
Gaussian fit routine {\it mpfit2dpeak}.  For each observation, the average
astrometric shifts of the bright sources in its FoV were used to define
the boresight shift for the observation.  We assume that there is no
offset between the two modules and only translational shifts are needed
for astrometric correction \citep{Harp10}. 
For five observations with no clear bright sources, we use 
the original coordinates
without any shifts. Table~\ref{t:obs} lists the applied boresight shifts
and the bright sources used for astrometric correction. 
Fig.~\ref{f:bs} shows the distribution of the boresight shifts, which
range from 3\arcsec\  to 11\arcsec.

The above approach implicitly assumes that the flux of bright sources
used for the boresight correction remained constant from observation to
observation: or additional iterations are needed to improve the accuracy
of astrometric correction. Since the astrometric errors of the final
source list based on the inital correction are well within the expected
performance of the \nustar optics
($<$5\arcsec\  positional errors for the 14 sources used for boresight
shifts: see \S\ref{s:catalog}), we did not perform further iterations.  


\begin{figure*} \begin{center}
\includegraphics*[width=0.97\textwidth,clip=true] {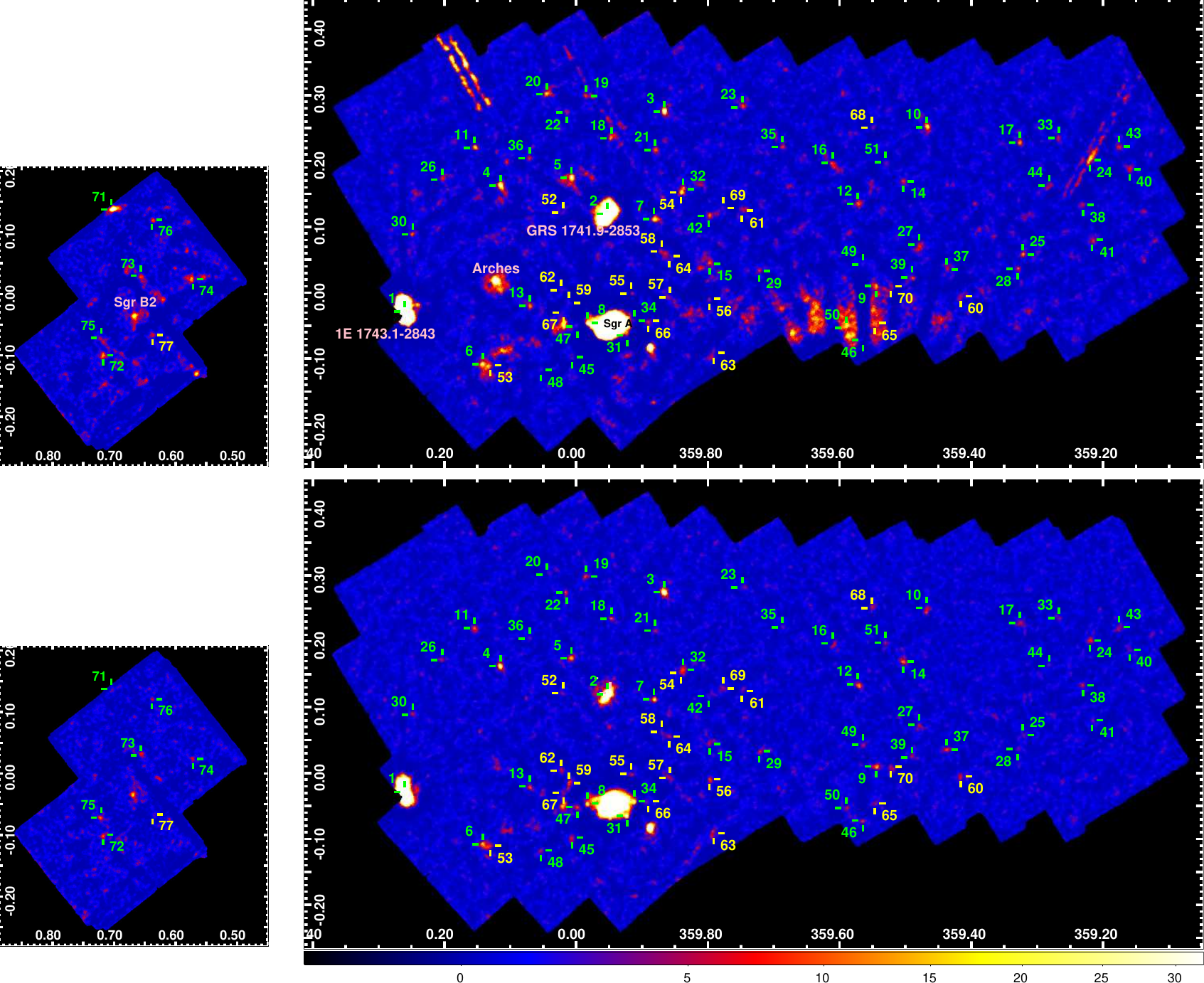}
\caption{Trial maps of the GC region in the 3--10 (top) and 10--40 keV
(bottom) bands using source cells of 20\% PSF enclosures,
overplotted with the \chandra counterparts of the \nustar detections
(green: group 1 and yellow: group 2, \S\ref{s:catalog}). 
The colors are scaled with the logarithmic values ($X$) of trial numbers
($10^X$), and the maximum is set at $X$=32
to make faint sources stand out more clearly.
A few large blobs of
high significance include the Sgr~A diffuse complex, GRS~1741.9--2853
(\S\ref{s:GRS1741}), 1E~1743.1--2843 (\S\ref{s:1E1743}) and
the Arches cluster (\S\ref{s:diffuse}). The large streaks in the 3--10 keV band
are (GR) backgrounds from bright sources near the region.  }
\label{f:trial}
\end{center}
\end{figure*}

For mosaicking, we re-projected the event files of each observation onto a common
tangent point in the sky and merged all of the observations together.
We also stacked the data sets of the two modules to maximize
photon statistics.
We generated a broadband (3--79 keV) image on the common sky grid
of the merged event file. For the matching global exposure map,
we mosaicked the individual exposure maps
by sampling and adding exposure values for every sky pixel in the
broadband image. The images mosaicked in this way tend to be free of
anomalous changes at the FoV boundaries of the individual observations
since it avoids rebinning the different sky grids of the
individual exposure maps.
We generated a set of the raw count
images in six energy bands on the common sky grid: 3--10, 10--40,
40--79, 10--20, 20--40 and 80--120 keV.  Since the 
\nustar optics have essentially no effective area above 80 keV, the
80--120 keV image is used for a sanity check of the systematic errors.
Fig.~\ref{f:raw} shows an exposure-corrected smoothed \nustar image 
in the 10--40 keV band and the vignetting-free exposure mosaic of the
main GC region and the Sgr B2 field.

\section{Source Detection} \label{s:det}

\subsection{Motivation for a New Source Search Technique} \label{s:motivation}

Source search routines such as {\it wavdetect} \citep{Freeman02} and {\it
wvdecomp\footnote{By A. Vikhlinin;
http://hea-www.harvard.edu/RD/zhtools/.}} have been very successful
in finding point sources from X-ray images taken by \chandra, \xmm and
other X-ray telescopes.  These techniques rely on the correlation between
the wavelet kernels and the local count distribution of X-ray images.
As researchers lower the detection thresholds of these techniques in
hopes of finding fainter sources, it becomes essential to independently
validate faint sources detected near the thresholds
\citep[e.g.~M09;][]{Hong12}.  An
independent validation also alleviates a somewhat unavoidable 
subjectivity inherent in threshold
setting \citep[][]{Townsley11}.  In short, negative values used
in wavelet analyses, although enabling efficient source detection,
introduce in essence a ``subtraction" procedure, which can be 
inadequate in characterizing the detection significance of
X-ray sources from non-negative counts following Poisson statistics.

The relative size of the \nustar FoV to the point spread function (PSF)
is much smaller than those of \chandra or \xmm. The
ratio of the  FoV ($\sim$13\arcmin) 
to the Half-Power Diameter (HPD, 58\arcsec) and FWHM (18\arcsec) of the PSF
in \nustar
is only about 13 and 40, respectively, whereas in \chandra the ratio 
exceeds 1000 (FoV$\sim$17.5\arcmin\ and  HPD $<$1\arcsec\ at the
aimpoint) for near on-axis sources.
Each \nustar observation often 
misses a large portion of the PSF of many sources. 
A point source in the mosaicked data often comprises 
a number of neighboring observations with partial PSF coverage, 
varying exposures and different vignetting effects.  
This, combined with relatively large \nustar backgrounds with 
complex patterns, further limits the utility of the
conventional techniques for source search in the mosaicked
\nustar
data. Except for several self-evident bright sources, all other sources
detected by the conventional techniques will have to be re-evaluated by
an independent measure.

\subsection{Trial Maps: New Detection Significance Images} \label{s:trialmap}

A rigorous probabilistic approach using Poisson statistics
is appropriate in describing the significance of source detection
in images of positive counts.  
For a given estimate of background counts, one can calculate 
the probability of acquiring more than the observed {\it total} counts
solely from a random fluctuation of the background. This
probability provides, in fact, a direct indicator of how likely or
unlikely it is to have a new source.
One of the key aspects of this probabilistic approach is in avoiding
subtraction used for handling the background. \cite{Weisskopf07}
and \citet{Kashyap10} independently calculated this probability ($P$),
which is described by a normalized incomplete
gamma function ($\gamma$) of the total observed counts and the background estimate:

\begin{eqnarray}
 	P(N>N^*|\lambda_S=0, \lambda_B) 
 			& = &  \sum_{N=N^*+1}^{\infty} \frac{e^{-\lambda_B} \lambda_B^N}{\Gamma(N^*+1)} \nonumber \\
 			& = &  \frac{\gamma(N^*+1,\lambda_B)}{\Gamma(N^*+1)} \nonumber \\
 	 		& = & \frac{1}{\Gamma(N^*+1)} \int^{\lambda_B}_0 e^{-t} t^{N^*} dt 
\end{eqnarray}
where $N^*$ is the observed total counts, $\lambda_B$ the expected mean
background counts, and $\lambda_S$ the expected mean source counts.
The condition, $\lambda_S=0$, ensures that the probability is calculated for
a random fluctuation from the background counts without any source.
M09 employed Eq.~1 to validate
faint sources detected by {\it wavdetect} and {\it wvdecomp}.

Our new source search technique uses the above random fluctuation
probability as a basis for source detection without relying on other
search tools: we calculate $P$ at every sky pixel in the mosaicked
images.  For a given sky pixel,
we first define a source detection cell using a circular region
around the pixel and a background cell using a surrounding annulus.
Then we estimate $\lambda_B$ from the counts in the background
cell scaled by the ratio of the exposure sum of the pixels in the source
and background cells. $N^*$ is simply the total counts in the source cell.
Then one can calculate the random chance probability 
at the sky pixel using Eq.~1.  We repeat the procedure for every
pixel in the image to create a map of the random chance probabilities.

We generate the random chance probability map using three fixed size source cells
with radii of 8.5\arcsec, 11.1\arcsec\ and 17.0\arcsec\  (corresponding to
10, 20 and 30\% enclosures of the PSFs, respectively) in seven energy bands (3--79,
3--10, 10--40, 40--79, 10--20, 20--40 and 80--120 keV). 
The inner radii of the corresponding background cells are 40\arcsec,
51\arcsec\ and 51\arcsec\ (60\%, 70\% and 70\% of the PSFs), respectively and the
outer radii are set to be
5/3rd inner radii.\footnote{This choice is made to allow the background cells
to be large enough for high photon statistics but not too far off the
source cells.  Note that source search using multiple scales makes the outcome
robust and insensitive to a particular set of the radius selection.}
Larger cells enable detections of faint sources
in a region relatively free of nearby X-ray emission,
while smaller cells enable detections of bright sources embedded in a region
of bright X-ray emission.

Unlike the X-ray images taken by \chandra, where both the size and shape of
the PSFs change significantly across the FoV as a function of the off-axis and
roll angles, in
\nustar the size of the PSF remains more or less constant although
the shape varies to some extent with the off-axis and roll angles \citep{Madsen15}.  This justifies
using fixed-size source and background cells across the field for source search
in the mosaicked \nustar images. In fact, even if the PSF size varies, using fixed-size
cells simply means that the resulting probability map retains the features
of the PSF shape.  In principle, using position-dependent, precise PSFs
for source cells allows deconvolution of the PSFs from the image through
iterations with forward modeling of the emission geometries.
The procedure can be applied to event lists instead of images.  If an event
list carries the sub-pixel information  (e.g.~enabled by dithering),
using the event list can improve source localization or identification
of small structures in the emission geometry.  On the other hand, 
using fixed-size cells on images enables a rapid calculation of random chance
probability maps through fast Fourier transformations
(FFTs).  In this paper, we calculate the probability maps using raw
count images instead of event lists and leave applications of varying PSFs
on the event lists for future studies.
See \S\ref{s:resolve} for the resolving power of the \nustar optics and
trial maps using fixed-size cells and \S\ref{s:diffuse}
for diffuse emission structures in trial maps.


\begin{figure} \begin{center}
\includegraphics*[width=0.45\textwidth,clip=true,trim=0 0 0.2cm 0]{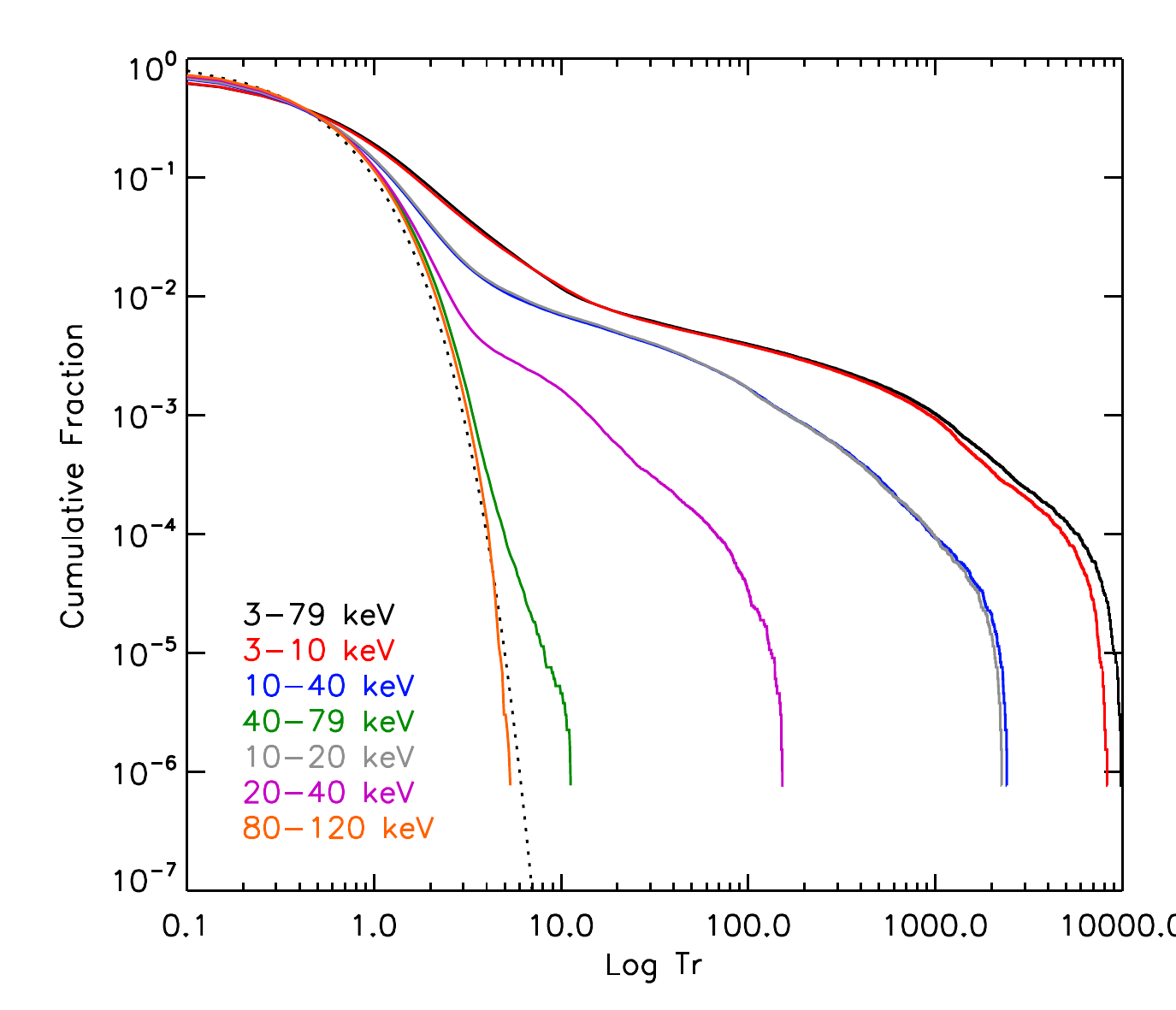}
\caption{Cumulative fractional distributions of the trial numbers (Tr=10$^X$)
in various energy bands with source cells of 20\% PSF enclosures.
Note the $x$-axis is effectively in a double logarithmic scale (i.e.~a logarithmic scale of $X$).
The observed distribution in the 80--120 keV band
matches with an ideal case of background-only random fluctuations
(the dotted line).  In the lower energy bands, the excess relative to
the ideal case is due to the observed celestial sources and the
associated systematics (e.g.~GR backgrounds).
}
\label{f:trialdist}
\end{center}
\end{figure}

Since the random chance probability ($P$) is opposite to the
probability of having a source (i.e., $1-P$), in order to mimic the usual sense of sky
images (i.e.~larger values for brighter sources with higher
significance), we use the inverse of
the random chance probability, which represents the number of random trials
needed to produce the observed counts by purely random fluctuations
of the background counts.  We call
the inverse of the random probability maps `trial' maps.
Fig.~\ref{f:trial} shows example trial maps generated in the
3--10 and 10--40 keV bands using 20\% PSF enclosures for source cells.
The colors are scaled with the logarithmic values ($X$) of the
required random trial numbers ($10^X$) and the maximum value of the
images is limited at $X$={32}  to make the faint sources stand out
more clearly.


\begin{figure*} \begin{center}
\includegraphics*[width=0.70\textwidth,clip=true]{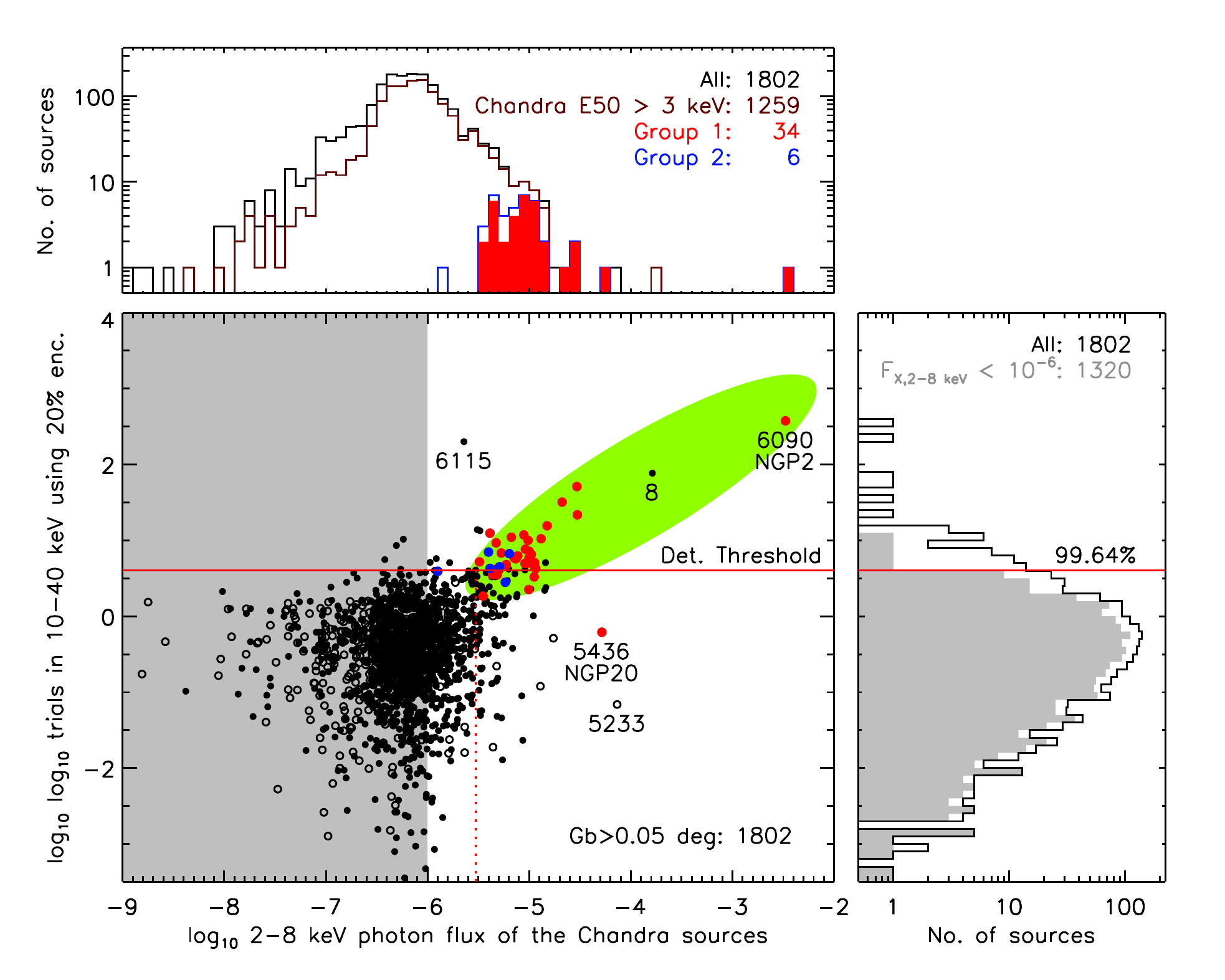}
\caption{
(Left) Scatter plot showing the \nustar 10--40 keV trial map values 
vs.~the \chandra 2--8 keV flux of the 1802 \chandra sources
at $G_B$ $>$0.05\Deg in M09.  The sources in the green ellipsoid
show a clear correlation, whereas the sources in the grey region at
$F_X$$<$ 10\sS{-6} \pcgs are uncorrelated.
The open symbols are likely foreground sources (with \chandra median
energies $<$ 3 keV). The \nustar detections are shown in red and blue for
group-1 and 2 sources, respectively (see \S\ref{s:catalog} for the definition
of the source groups). Sources above the threshold (the solid red line) or
with a relatively high \chandra flux ($>$3$\times$10\sS{-6} \pcgs, the
dotted red line) were all visually inspected for detection.
(Right) Double logarithmic distributions of the trial map values. 
The distribution of the sources with $F_X$$<$ 10\sS{-6} \pcgs 
is used for setting the detection threshold.
(Top) Distribution of the \chandra  2--8 keV fluxes.
\chandra source IDs \#8 and \#6115 are in the bright PSF wings of
GRS 1741.9-2853 (see \S\ref{s:resolve}). \chandra ID \#5436 (GRO J1744-28)
is detected only in the 3--10 and 3--79 keV bands (\S\ref{s:gro1744}).
\chandra ID \#5233, which is a foreground star, was not detected by
\nustar (see \S\ref{s:missing}).
}
\label{f:cor}
\end{center}
\end{figure*}

\subsection{Threshold Setting for Trial Maps} \label{s:thresh}

Trial maps provide the statistical significance of potential
sources, but the systematic errors need to be taken into account in
order to set a proper detection threshold and thus efficiently detect real
X-ray sources while minimizing false detections.
Fig.~\ref{f:trialdist} shows the cumulative fractional distributions
of the random trial numbers with source cells of 20\% PSF enclosures in
various energy bands.  The distribution in the 80--120 keV band, where
the \nustar optics has no response to incoming X-rays, is consistent with
an ideal case of purely Poisson statistics-driven random fluctuations
of uniform backgrounds (dotted
line).\footnote{The cumulative fraction distribution for the ideal
background-only case is simply an inverse function of the trial numbers.
i.e.~10\sS{-X}.} The match indirectly 
indicates that there
are no apparent systematic errors in the detector system or in 
the data processing including the mosaicking  procedure.
The large
excess in the lower energy bands relative
to the ideal case originates from the observed celestial sources and
the associated systematics of the X-ray optics (e.g.~GR backgrounds).

For a given trial map, a statistically conservative detection threshold
can be simply the number of pixels ($\sim$\nep{5}{5} for the main GC region) in
the map under the assumptions that each pixel represents an independent
search attempt and that one false detection is allowed over the entire
map.  Since source cells used for search are much larger than a pixel
(e.g.~36 pixels in a detection cell of the 15\% PSF enclosure), the actual
number of independent search attempts in the map is much smaller than
the number of pixels. Therefore, the pixel-count based threshold can
be a conservative limit for source search in the trial maps
of high energy bands ($>$40 keV) where the statistical errors dominate
the systematic errors. 

For the trial maps of low energy bands below
40 keV, the pixel-count based threshold
is still not stringent enough due to the large
systematic errors as seen in Figs.~\ref{f:trial} and \ref{f:trialdist}.
The main cause of the systematic errors in trial maps is the inaccuracy in
estimating the true mean background counts ($\lambda_B$) in Eq.~1. We use
the scaled counts of background cells for $\lambda_B$, but the background
is not uniform. In particular, the contamination from the residual SL and
GR backgrounds or large PSF wings of bright neighboring sources does not
scale simply by the exposure ratios between the source and background
cells.  In principle, these systematics can be forward-modeled after
initial detections, which would require extensive simulation and modeling work
due to the diverse geometries and spectral types of the diffuse and point
sources in the GC region.  Instead, we evaluate the contribution of the
systematics in the trial maps using a deep \chandra source catalog by
M09 and set proper detection thresholds accordingly.

First, we exclude the regions clearly contaminated by the PSF wings
of bright diffuse and point sources.  Then, we cross-correlate the
remaining region of each trial map with the \chandra source catalog.
Except for highly variable sources, we expect that the majority of the
\nustar sources have \chandra counterparts, so we first search for
the \nustar detection of the \chandra sources.
Fig.~\ref{f:cor} shows a scatter plot of 
the \chandra 2--8 keV fluxes of the \chandra sources in the GC region M09
and the \nustar 10--40 keV trial map values at the \chandra source positions.
For easy illustration, we only show the sources at
Galactic latitudes $G_B$$\ge$0.05\Deg, where no bright diffuse
features are observed in
the \nustar 3--79 keV band. Evident is the correlation between the bright
\chandra sources and their \nustar trial numbers as highlighted by a
green ellipsoid, whereas the sources lying in the grey region at
$F_{X}$$<$10\sS{-6} \pcgs in the 2--8 keV band are uncorrelated.  For threshold
setting, we generate a subset of the trial number distribution using these
uncorrelated sources as shown in the shaded histogram on the right panel.

We search for sources in the 18 trial maps (the six energy bands below 80
keV and the three cell sizes).  The 18 trial maps are independent of each
other in varying degrees. For example, the 3--10 and 10--40 keV trial
maps are generated completely independently, while the 3--79 and 10--40 keV trial
maps share some common data.  We only consider a source as valid in
the final list if the source is found to be above the threshold in at least
two trial maps.  For simplicity, we assume that all the trial maps
are independent of each other.  Then if we require a certain percentage
($p$) of the false sources to be rejected in each map, the expected
false sources ($N_F$) in the final list is calculated as $N\Ss{can}\
C(18,2)\  p\sS{16} (1-p)^2$ where $N\Ss{can}$ is the number of 
\chandra sources to consider in search for the \nustar detection and $C(i,j)$
is combinatorial or binomial coefficient.  To account for some dependency
between the maps, we put a tight limit on $N_F$  by setting it at 0.5
instead of 1.

Judging from the correlation pattern in Fig.~\ref{f:cor},
we search the \nustar detection of the \chandra sources only with
$F_X$$\ge$3\x10\sS{-6} \pcgs in 2--8 keV. 
In the main GC region, we have $N\Ss{can}$=264, and the required
rejection percentage ($p$) for $N_F$=0.5 is 99.64\%.  The
corresponding thresholds range
from 10\sS{2.7} in 40--79 keV with source cells of 15\% PSF
enclosures to 10\sS{10.2} in the 3--79 keV with 30\% PSF
enclosures.  We also use $N\Ss{can}$=$N\Ss{pix}$ to calculate a conservative
lower limit of the thresholds common for all the maps, which
is 10\sS{4.1}. 

For initial screening, we consider all the sources above the thresholds
(regardless of their positions, without any exclusion zone\footnote{The
exclusion zone (e.g.~the large diffuse complex) was only used for setting
the thresholds.}) and all the sources
with $F_X$$>$3$\times$10\sS{-6} \pcgs in 2--8 keV (regardless of their
\nustar trial numbers). They add up to 290
sources in the main GC region.  We visually inspect these 290 candidates
in the 18 trial maps and their position in the 18 scatter plots
similar to Fig.~\ref{f:cor}. The number of the
initial candidates is large because of many \chandra sources located in the
large diffuse complex near Sgr~A*.  In the final list we exclude the
sources in bright diffuse structures if they are not clearly resolved.

The sources in the Sgr~B2 region were selected from their own set of the
thresholds by the same procedure. In order to minimize false detections
arising from the bright SL background in the Sgr B2 field, we repeat the procedure with two
different SL cuts, and only the detections that are common in both cases
are selected as real sources.

We also search for \nustar-only sources without matches to \chandra
counterparts or possibly missed detections due to the relatively large
positional uncertainty of the \nustar optics (18\arcsec\ FWHM). 
To do so, we look for any
spots above the thresholds in more than two trial maps outside of the
10\arcsec\ radius of the \chandra positions of the \nustar detections.
We have found two such sources,\footnote{When
searching for \chandra sources in \nustar trial maps, the values
are sampled {\it at} the \chandra source positions: i.e.~these two \chandra
sources have sub-threshold trial values at their \chandra source
positions, which is the reason that they were not selected in the
original search. This approach was chosen to simplify the search
procedure in comparison to an alternative method that allows some
positional uncertainty in the original search.}
NGPs 61 and 68.  Given the high density
of the \chandra source population in the region, both of the detections
have a \chandra source within the positional uncertainty of the \nustar
optics, which is assigned as a potential counterpart.

\subsection{Source Catalog} \label{s:catalog}


\begin{figure} \begin{center}
\includegraphics*[width=0.45\textwidth,clip=true]{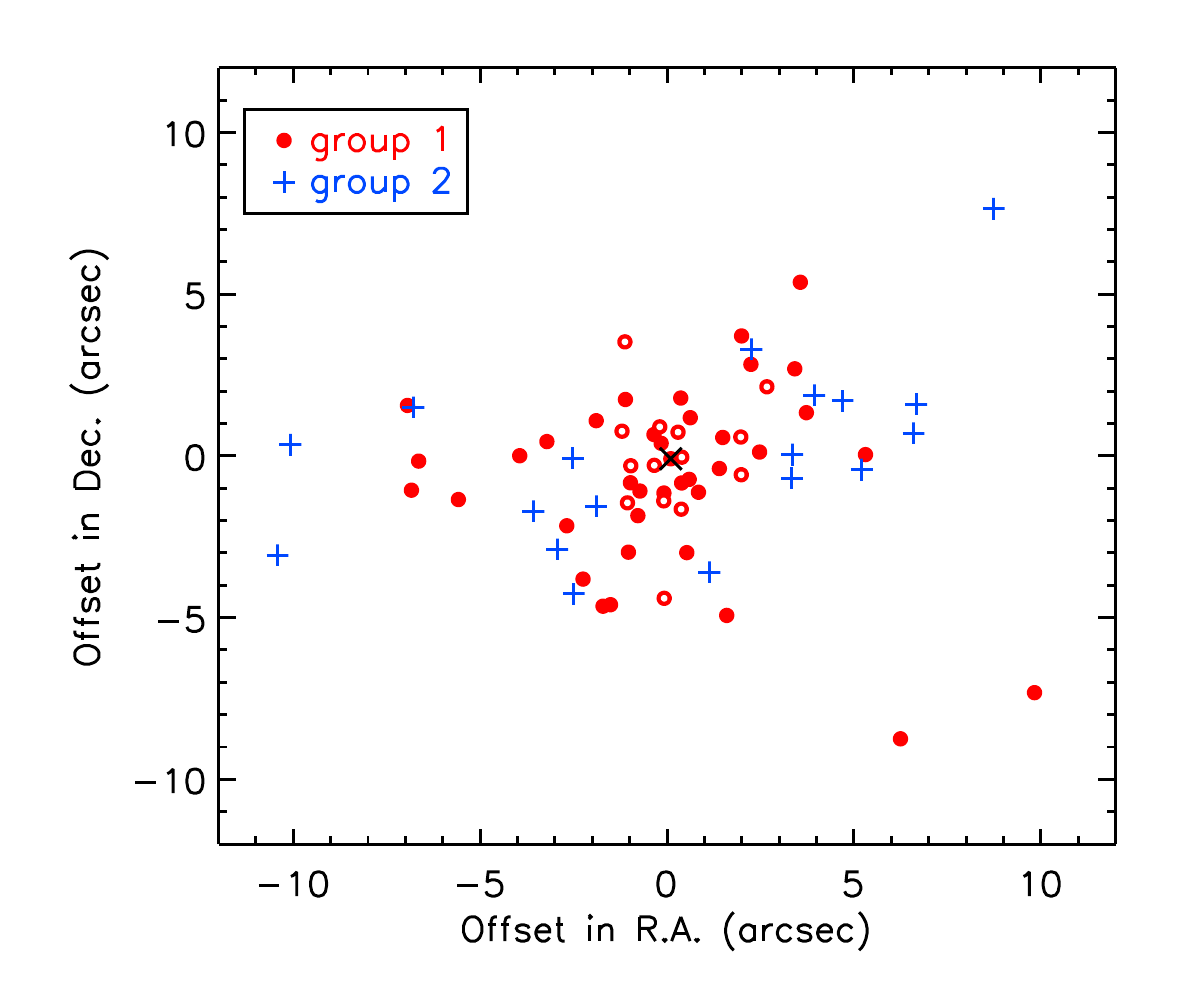}
\caption{Offset distribution between the \nustar and \chandra positions
in the main GC region.
The open circles indicate the sources used for astrometric correction.
The median offset ($\sim$0.1\arcsec) of the distribution is marked by
an `x' symbol.
}
\label{f:offset}
\end{center}
\end{figure}

Tables~\ref{t:src} and \ref{t:src:SgrB2} show the final source catalog
of the main GC region and the Sgr~B2 region, respectively.  Each table
divides the sources into two groups and they are listed in decreasing order
of the maximum trial value of the 18 maps.  
The columns of the tables are as follows.

\begin{compactenum}
\item \nustar Galactic Center Point (NGP) source ID.
\item The local peak location of the trial map within 30\arcsec\ of the
\chandra position. They are weighted average values among the trial maps with detections.
For sources with bright neighbors, we limit
the search radius to 10\arcsec\ or 15\arcsec\, depending on the proximity.
The peak position is determined by a 2-D Gaussian fit on the trial map.
\item (Candidate) \chandra counterpart ID by M09.
\item (Candidate) \chandra counterpart name. 
\item The \chandra 2--8 keV flux of the counterpart.
\item The angular offset between the \nustar and \chandra positions. 
\item The combined exposure of the two \nustar FPMs at the
\chandra source positions.
\item An indicator of the soft (S, $<$ 10 keV) and/or hard (H, $>$ 10 keV)
band detection.
\item The trial map value at the \chandra position. 
The sources are ordered by this value. 
\item The energy band of the trial map with the local peak value.
\item The source cell size of the trial map with the local peak value. 
\item The number of trial maps above their respective thresholds at the \chandra positions.
\item A known name, nearby \chandra source, and/or notable diffuse feature nearby.
\end{compactenum}

The sources in group 1
have a relatively clear \chandra counterpart which is
usually the only bright ($F_X$$>$3$\times$10\sS{-6} \pcgs) \chandra
source around the \nustar detection.  The sources in group 2
have solid \nustar detections (except for NGP 65, which is a bit marginal),
but their association with the \chandra sources is not as
clear either because multiple \chandra sources of similar fluxes are
found within the uncertainty of the \nustar positions (e.g.~NGPs 55 and 56) or
because a diffuse origin of the hard X-ray emission cannot be ruled out
(e.g.~NGPs 53 and 59, see \S\ref{s:hardsources}).  After visual
inspection of all the \nustar detections, we have 58 group-1 and 19
group-2 sources.  

Fig.~\ref{f:offset} illustrates the offset distribution between
the \nustar and \chandra positions of the \nustar detections in the
main GC region.
The median offset of the distribution is $\sim$0.1\arcsec, and the
distribution does not show any significant systematic offsets,
validating the astrometric correction of the individual observations.
The sources in group 1 show relatively smaller offsets than in group 2,
which is in part because group 1 includes the 14 bright sources used for
astrometric correction.  The maximum offset is 12.3\arcsec\  for NGP
27.  Of five sources with more than 10\arcsec\ offsets,
two sources (NGPs 61 and 68) are found during the search for
\nustar-only detections (\S\ref{s:thresh}); neighboring X-ray
emission or nearby artifacts likely contributed to the large offsets
of the other three.

\section{Aperture Photometry} \label{s:ap}

\subsection{Aperture Selection for Photometry} \label{s:aperture}

For photometry, we use a circular region and an annulus centered
around the \chandra position of each source as a basis for source and
background apertures, respectively.  These apertures are similar to 
the detection cells used for the source search, 
but the former are usually chosen to be larger than the latter
in order to attain higher
photon statistics needed for estimation of various source properties.
For example, detection cells used in {\it wavdetect} 
for source search in \chandra X-ray images typically start with about 30--40\%
PSF enclosure circles \citep{Freeman02}, whereas 
apertures for photometry are typically about 80--95\% PSF enclosure
circles \citep{Broos10}.


\begin{figure} \begin{center}
\includegraphics*[width=0.475\textwidth]{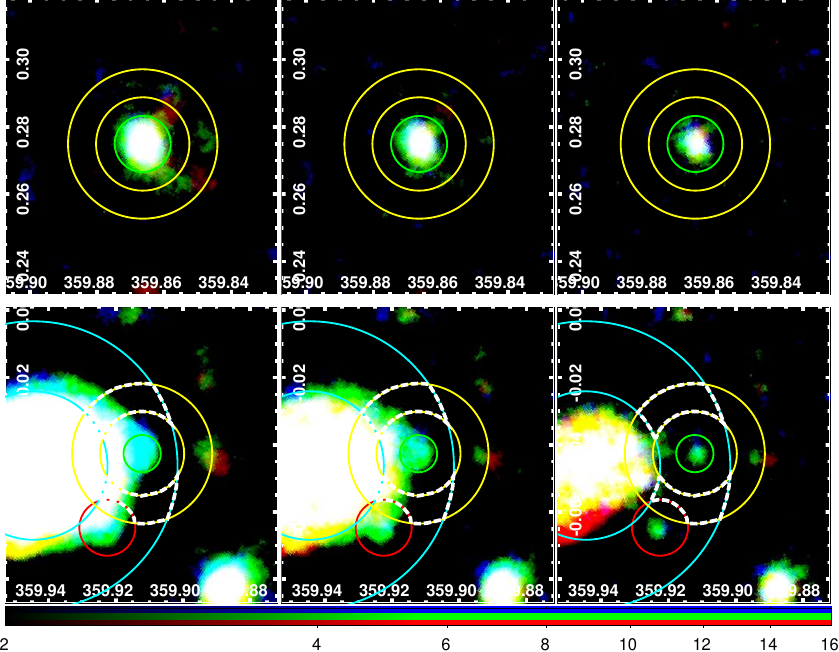}
\caption{Example aperture selections for photometry overlaid in the
trial maps centered around NGP 3 (top) and NGP 34 (bottom). 
From left to right, 
the trial maps from 30\%, 20\% and 15\% PSF enclosures for
source cells are shown to illustrate the scales of the source 
and surrounding diffuse emission relative to the aperture selections.  
In the case of NGP 34, the intersection of the two
annuli (yellow and cyan) excluding the neighboring source (red)
is used for the background aperture (the dashed lines).
The colors are scaled with the logarithmic values ($X$) of trial
numbers ($10^X$).
}
\label{f:example}
\end{center}
\end{figure}

Using apertures symmetric with respect to the source position tends to
be effective in alleviating artifacts in the X-ray optics and detector
response and also in eliminating the internal and external background
components of low spatial frequencies.
For instance, the trial maps (Fig.~\ref{f:trial})
generated with the symmetric detection cells lack 
the large scale diffuse emission that is evident in
the (smoothed) raw images (Fig.~\ref{f:raw}).

We use two baseline sets of aperture sizes
to assess the systematic errors intrinsic to aperture selection. The
first set uses 30\arcsec\ radius circles ($\sim$50\% PSF enclosures)
for source apertures and annuli of 50\arcsec\ to 80\arcsec\ radii
for the matching background apertures. The second set uses 40\arcsec\
radius circles ($\sim$60\% PSF enclosures) for source and 60\arcsec\
to 90\arcsec\ annuli for background.
The baseline apertures work
well for relatively isolated sources (about 65\%: e.g.~top panel in
Fig.~\ref{f:example}), judging from the fact that the estimates of the 
absorbed photon fluxes in the 3--10 keV band are consistent\footnote{Note that there is a small
calibration discrepancy (about 15\% level) between the \chandra and
\nustar responses \citep{Madsen15}. In addition, depending on the source spectrum, the
difference of the two energy ranges will introduce a small difference
in the flux estimate (e.g.~about 14\% for an absorbed power-law model
with $\Gamma$ = 1 and
\nH = 6$\times$10\sS{22} cm\sS{-2}). However, these differences are at
the level of the 1$\sigma$ error of the \nustar flux estimate except
for the first two brightest sources.} with the reported 2--8 keV
\chandra fluxes in M09 within 3$\sigma$.

The remaining sources with bright neighbors require additional care
in aperture selection.  For these sources, the photometry results are
often too sensitive to the size of the background apertures.
 For instance, for NGPs 31 and 34 that are
located near the edge of the diffuse emission complex around Sgr~A*,
the gradient of the emission structure plays an important role in the
photometry results.  To make aperture selection more objective and thus
aperture photometry more reliable, we assume that the soft X-ray fluxes
below 10 keV of these sources have not changed
significantly from the \chandra fluxes reported by M09. 
Under this assumption,
first we reduce the radius of the source aperture by 10\arcsec\ to
limit the contamination.  Then we gradually exclude parts of the
background aperture that are somewhat dominated by the X-ray emission
of the neighbors while
maintaining the symmetry of the aperture shape as much as possible
until we get an agreement in photon fluxes between the
\nustar 3--10 keV and \chandra 2--8 keV bands within a factor of
few.\footnote{It is not unusual to observe a flux variation by a
factor of few from a faint source with a constant luminosity when the
observed photon statistics are poor.  See \S\ref{s:var}.}

Fig.~\ref{f:example} shows an example aperture selection of NGP 34.
We exclude the emission from the Sgr~A diffuse complex in the background
annulus (yellow) using another annulus (cyan) centered around Sgr~A*:
we use the intersection of the two annuli for the background aperture.
We also exclude the contribution from the neighbor NGP 31 
(red).  These modifications,  although a bit ad hoc, retain
the benefits of having (more or less) symmetric apertures and enable
a somewhat consistent scheme in aperture selection for all the sources.  
For bright sources with large PSF wings (e.g.~GRS~1741.9--2853), we extended
the source and background apertures accordingly. 

\subsection{Photometry Results} \label{s:photometry}

For each source, we extract the events in the source and background
apertures from the merged event file and calculate the net counts
for a set of energy bands.  The relative scale between the source and
background apertures is given by the ratio of the summed exposure values
(no-vignetting) of the two apertures.
Table~\ref{t:ap} lists the photometry results.  The columns are defined
below and the next few sections describe how we estimate some of
the source properties in the table.

\begin{compactenum}
\item \nustar Galactic Center Point (NGP) source ID.
\item \chandra source ID by M09.
\item The net counts in the 3--40 keV band.
\item The mode of the posterior distribution of Bayesian Enhanced
X-ray Hardness Ratio \citep[BEHR:][see \S\ref{s:class}]{Park06}: ($H-S$)/($H+S$)
where $H$ and $S$ are net counts in 3--10 and 10--40 keV, respectively.
\item The median energy of the \nustar spectrum in 3--40 keV.
\item A relative ratio of 25\% and 75\% quartiles: 3 ($E_{25}$ -- 3 keV)/($E_{75}$ -- 3 keV),
equivalent to the $y$-axis value in the \nustar quantile diagram \citep[][see \S\ref{s:class}]{Hong04}.
\item An estimate of \nH along the line of sight \citep{Nishiyama08}.
\item An estimate of photon index using the median energy for an absorbed
power-law model with \nH = 6$\times$10\sS{22} cm\sS{-2} (\S\ref{s:class}).
\item The observed (i.e.~absorbed\footnote{X-ray photon fluxes and luminosities
quoted in this paper are all absorbed quantities using the assumed or
estimated \nH values unless otherwise noted.})
\chandra 2--8 keV flux from M09.
\item The observed \nustar 3--10 keV flux (\S\ref{s:flux}).
\item The observed \nustar 10--40 keV flux (\S\ref{s:flux}).
\item The observed \nustar 3--10 keV luminosity at 8 kpc (\S\ref{s:flux}).
\item The observed \nustar 10--40 keV luminosity at 8 kpc (\S\ref{s:flux}).
\item The source and background aperture radii.
($a$) 20\arcsec/35\arcsec--42\arcsec, 
($b$) 20\arcsec/30\arcsec--46\arcsec, 
($c$) 20\arcsec/45\arcsec--75\arcsec, 
($d$) 30\arcsec/45\arcsec--45\arcsec, 
($e$) 30\arcsec/50\arcsec--80\arcsec, 
($f$) 40\arcsec/60\arcsec--90\arcsec, 
($g$) 8\arcsec/130\arcsec--145\arcsec, 
($h$) 70\arcsec/145\arcsec--145\arcsec, 
($i$) 70\arcsec/210\arcsec--230\arcsec, 
($j$) 100\arcsec/210\arcsec--230\arcsec.
We performed aperture photometry using two aperture sets for each source:
the first set to provide the basic photometry results,
and the second set to estimate the systematic errors originating
from the aperture selection (\S\ref{s:aperture}).
The two baseline choices are ($e$) and ($f$).
An underline indicates the aperture has additional exclusion zones
(see \S\ref{s:aperture}). 
\item 
The \nustar flags:
($f$) sources showing the iron lines (\S\ref{s:fit});
($k$) sources with short-term variability according to a KS test of individual
observations (\S\ref{s:var});
($r$) sources with long-term variablility according to the
maximum-to-minimum flux ratio of multiple observations (\S\ref{s:var}).  
The \chandra source flags (M09): 
($c$) sources confused with another nearby source;
($g$) sources that fell near the edge of a detector 
	in one or more observations; 
($b$) sources for which the source and background spectra have a
	$>$10\% chance of being drawn from the same distribution according to
	the Kolmogorov-Smirnov (KS) tests; 
($s$) sources variable on short time scales, as
	indicated by probabilities of $<$1\% that the event arrival times for
	at least one observation were consistent with
	a uniform distribution according to the KS test; 
($l$) sources that were variable on long time scales, as indicated
	by a probability of $<$1\% that the fluxes for all observations were
	consistent with a uniform distribution according to the KS test. 
Others: ($t$) transients identified in \citet{Degenaar12}.
\end{compactenum}

The errors quoted in Table~\ref{t:ap} are the largest of the three
estimates: a statistical error and two different estimates of systematic
errors. The statistical error is estimated from the uncertainty of the
observed net counts after background subtraction.  A systematic error is
given by the difference in the photometry results between two aperture
sets (marked with *).  In calculating the photon indices and the X-ray
luminosities, another systematic error is estimated based on the
selection of spectral model parameters (marked with $\dagger$, see
\S\ref{s:flux}).


\begin{figure*} \begin{center}
\includegraphics*[width=0.99\textwidth,clip=true]{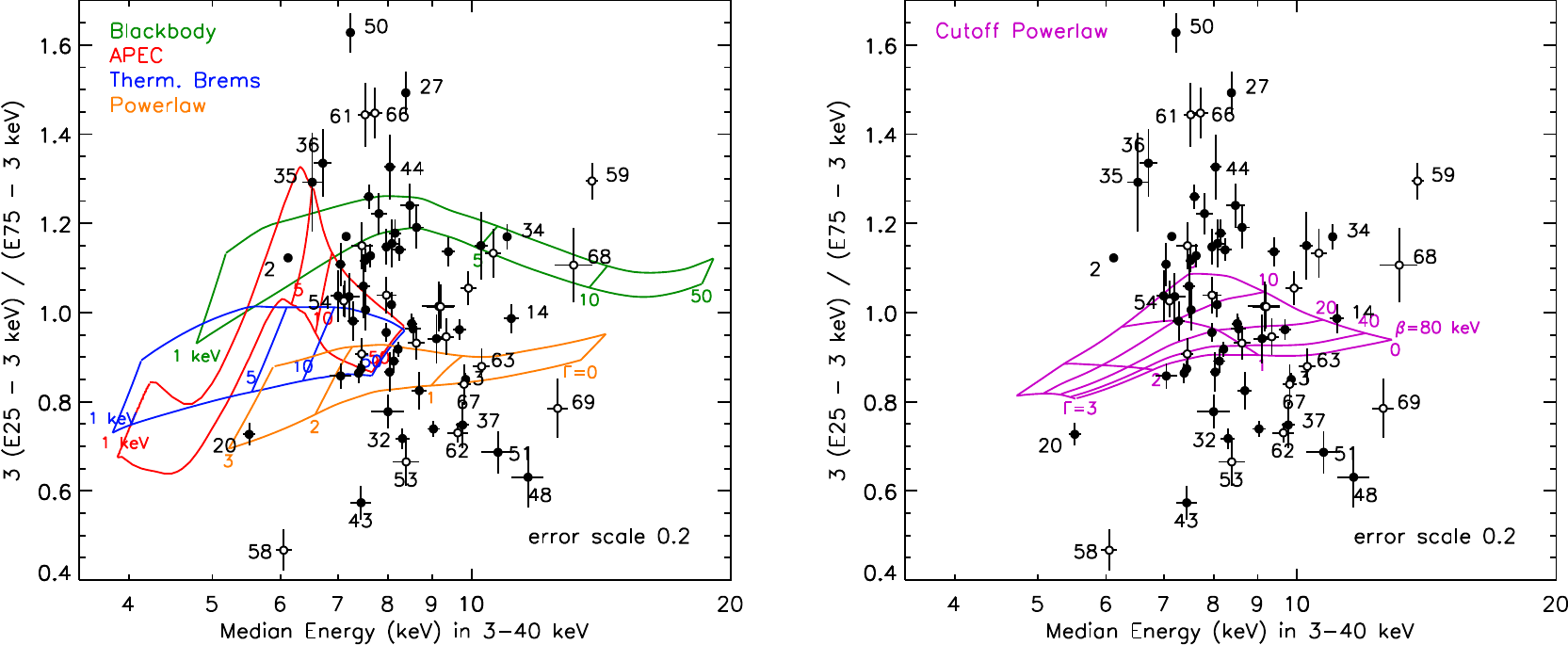}
\caption{Quantile diagrams of the \nustar sources in comparison with
five spectral models.  The grids on the left panel are for absorbed
power-law (yellow, from right to left, photon indices of 0, 1, 2, and 3),
thermal bremsstrahlung (blue), APEC (red), and blackbody (green) models.
The thermal models cover $kT$ of 1, 5, 10 and 50 keV, which run from
left to right.  The height of the grid pattern in each model represents
the variation between \nH = 10\sS{22} and 10\sS{23} cm\sS{-2}.  
The grid on the right panel is for an absorbed cut-off power-law model
[$E^{-\Gamma} \exp(-E/\beta)$] with \nH = \nep{6}{22} cm\sS{-2}. The
covered photon indices ($\Gamma$) are 0, 1, 2 and 3, and the cut-off
energies ($\beta$) are 5, 10, 20, 40 and 80 keV.
The closed and open circles are from the group-1 and 2 sources,
respectively. The error bars are scaled down to 20\% of the original values
for easy viewing.
}
\label{f:qd}
\end{center}
\end{figure*}

\subsection{X-ray Hardness Ratio and Energy Quantiles} \label{s:class}

We use the Bayesian Estimation of Hardness Ratios \citep[BEHRs;][]{Park06} 
and the energy quantiles \citep{Hong04} to classify the spectral types
of the \nustar sources.  Conventional hardness ratios or X-ray colors
are often subject to a spectral bias intrinsic to the choice of the 
energy bands.  The BEHRs alleviate the intrinsic spectral bias through
a more rigorous probabilistic approach. Energy quantiles are free of such
a spectral bias and enable an easy classification of diverse spectral types.

We use the BEHR between the 3--10 and 10--40 keV bands and the median
energy in 3--40 keV as an illustrator of the overall spectral hardness.
The quoted value of the BEHR is the mode of the posterior distribution
of $(H-S)/(H+S)$ where $H$ and $S$ are net counts in 3--10 and 10--40
keV bands, respectively.  The error represents the larger deviation of
the $\pm$34\% range (1$\sigma$ equivalent) of the posterior distribution.
The error of an energy quantile is given by the standard deviation of
the quantiles from 100 randomly selected half-sampled sets of the source
events.

For a two-parameter classification, X-ray color-color diagrams are often
used, but the poor statistics and the diverse spectral types frequently
result in only upper or lower limits for many estimates of X-ray colors.
We use quantile diagrams consisting of the median
energies vs.~the quartile ratios (see also \S\ref{s:missing}). 
Fig.~\ref{f:qd} shows \nustar quantile diagrams in 3--40 keV overlaid
with several spectral model grids.  The grids on the left panel
indicate power-law (yellow), thermal bremsstrahlung (blue), APEC (red)
and blackbody
(green) models with absorptions of \nH = 10\sS{22} and 10\sS{23} cm\sS{-2}
to guide the spectral type of the \nustar sources.  
We use \citet{Anders89} for the abundance model in the absorption.
The power-law model
covers $\Gamma$ = 0, 1, 2 and 3, and the thermal models cover $kT$
= 1, 5, 10 and 50 keVs.
The grid on the right panel is for an absorbed cut-off power-law model [$E^{-\Gamma} \exp(-E/\beta)$]
with \nH = \nep{6}{22} cm\sS{-2}. The cut-off energies ($\beta$) cover 5, 10, 20, 40 and 80 keV,
and the photon indices ($\Gamma$) range 0, 1, 2, and 3.
The error bars are scaled down
to 20\% of the original values for easy viewing.

The quantile diagram illustrates that the X-ray spectra of some sources
(e.g.~NGP 2) are better described by thermal plasma models while others
(e.g.~NGP 3) by non-thermal power-law models, which is not clear
from the BEHRs or the median energies alone.  Sources that lie in between
the blackbody and power-law model grids can be better described by an
absorbed cut-off power-law model as seen on the right panel.  However,
due to the relatively large uncertainties of the quartile ratios and
the degeneracy among the different spectral models in the diagram, it
is not straightforward to assign both spectral models and parameters
for many sources from the quantile diagram.

\subsection{Flux and Luminosity Estimates} \label{s:flux}

To properly account for the spectral type in estimation of observed photon flux
and luminosity of each source, we use an absorbed power-law model with
the median energy-based photon index.
We also assume $\nH$ = 6$\times$10\sS{22}
cm\sS{-2}, which is often used to describe the interstellar
absorption for sources in the GC region (e.g.~M09).
For comparison, Table~\ref{t:ap} also lists an estimate along the line of
sight towards each source based on the $A_{Ks}$ map of the GC region generated
from the observations by the SIRIUS camera on the Infrared
Survey Facility telescope \citep{Nishiyama08}. 
The resolution of the $A_{Ks}$ map is 15\arcsec.
Their values range from \nH $\sim$ 3--7$\times$10\sS{22}
cm\sS{-2}, whereas the estimates based on the \nustar quantile diagram
(or the X-ray color-color diagrams) are often higher with large
uncertainties. Some of the bright sources such as the Cannonball (NGP
8) indeed exhibit an X-ray spectrum with a higher extinction than
the field estimate or the assumed value, which may be the result of the
local absorption around the source, but the \nustar X-ray band ($>$ 3 keV)
is not sensitive to the absorption below \nH $\sim$ 10\sS{23}
cm\sS{-2}  (\S\ref{s:cannonball}).

\begin{figure*} \begin{center}
\includegraphics*[width=0.99\textwidth,clip=true]{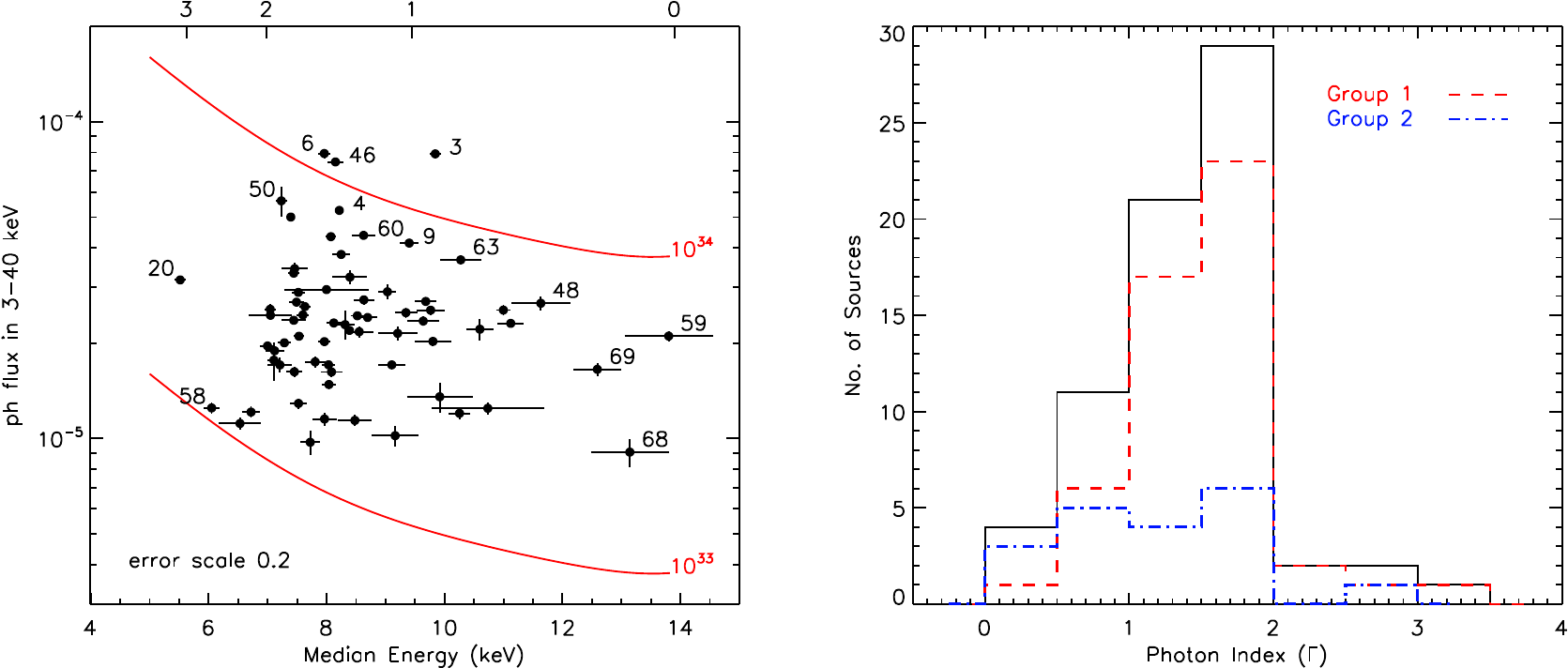}
\caption{Scatter plot of photon flux vs.~median energy 
of the \nustar sources (left) and their equivalent photon index
distribution (right).
The top $x$-axis on the left panel shows the equivalent photon indices for an absorbed power-law model with
\nH = 6$\times$10\sS{22} cm\sS{-2} (\S\ref{s:class}). The two lines show the
iso-luminosity tracks for $L_X$ = 10\sS{33} and 10\sS{34} \lcgs
at 8 kpc in the 3--40 keV band. The error bars are scaled down to 20\% of the original values
for easy viewing.
The red, blue and black histograms shows group-1, 2 sources and the sum of the two,
respectively.  }
\label{f:dist}
\end{center}
\end{figure*}

To assess the systematic error arising from an inaccurate assumption of the
extinction, we re-estimate the photon index by changing the $\nH$
value to 3\x\ and 12$\times$10\sS{22} cm\sS{-2}. We also
re-calculate the photon index by varying the median energy
by 1$\sigma$ with the $\nH$ value fixed at 6$\times$10\sS{22}
cm\sS{-2}. The systematic error is given by the largest difference
between the original estimate and these four estimates.  
This systematic error is quoted with $\dagger$ in Table~\ref{t:ap},
if it is larger than the statistical error and 
the difference between the two aperture sets (\S\ref{s:aperture}).

The left panel in Fig.~\ref{f:dist} shows a scatter plot of absorbed
3--40 keV photon flux vs median energy of the \nustar sources.
The tracks show the iso-luminosity
lines for absorbed power-law models with \nH = 6$\times$10\sS{22}
cm\sS{-2}.  The top-axis shows the equivalent photon indices.
The error bars on the left panel are scaled down
to 20\% of the original values for easy viewing. 
The right panel in Fig.~\ref{f:dist} shows a distribution
of equivalent photon index of the \nustar sources for an absorbed
power-law model. 

Table~\ref{t:ap} shows the observed photon fluxes calculated
for an absorbed power-law model with the median energy-based photon
index and \nH = 6$\times$10\sS{22} cm\sS{-2}. We also estimate the photon
fluxes non-parametrically (not shown in the table), where we
calculate the net counts in every 1 keV step,
convert them to the matching photon fluxes by using the
Auxiliary Response Function (ARF) of the source 
and then summing them over a given energy band. 
This direct conversion from photon counts to fluxes is not usually
encouraged because the conversion is prone to large amplification of 
statistical noise.  On the other hand, the non-parametric estimation
offers a sanity check of the model dependence in the flux and
luminosity estimations (see \S\ref{s:lnls}).
The difference between the model-based and model-independent estimates
are less than 40\% except for a few of the faintest sources. On average, the
non-parametric estimation is about 6\% to 11\% lower than the
power-law model-based estimation, depending on the energy bands.

For estimation of observed luminosities, we use an absorbed power-law model
and a distance of 8 kpc for all the sources with \nH = \nep{6}{22} cm\sS{-2},
assuming they all are in the central Galactic
Bulge (\S\ref{s:missing}).  The uncertainty in \nH is not a dominant
factor of uncertainties in the flux and luminosity estimations. 
For several brightest sources, we compare the estimates from
the spectral model fits with simple median-energy based estimates in
\S\ref{s:bright}.

\subsection{Spectral Model Fit for Bright Sources} \label{s:fit}

For some of the bright \nustar sources with net counts greater than $\sim$600
(excluding ones already in the literature),
the spectra were analyzed through spectral model fitting in addition
to the spectral classification described in \S\ref{s:class}.  We also
search for the \chandra and \xmm archival data, and if available, we
jointly fit the \nustar spectra with the \chandra and/or \xmm spectra.
The \chandra spectra are from M09.  The \xmm spectra are from the \xmm
pipeline processing system. For sources with multiple \xmm observations,
we regenerate a spectrum of each observation and stack them together to
get a combined spectrum.

We generate a combined \nustar spectrum for each source by stacking
individual X-ray spectra from multiple observations with proper
scalings using the FTOOL {\it addspec}.  As aforementioned, if a source is
covered by multiple observations, it is bound to fall near the edge of a
chip in some of them.  Those observations that miss a large portion of the
PSF are excluded in building the stacked spectrum since their individual
spectra are of poor statistics and their instrumental responses are likely
subject to a large uncertainty.  As a result, the stacked X-ray spectra
of many sources do not have sufficient photon counts to put meaningful
constraints on the spectral parameters through model fitting.  In other words,
high detection significance in Table~\ref{t:src} or high net count
in Table~\ref{t:ap} does not guarantee a \nustar X-ray spectrum with
high signal-to-noise ratio.  Out of the nine \nustar sources with net
counts greater 600, Table~\ref{t:src:fit} summarizes the best-fit
parameters of four sources with
relatively good spectral fits for absorbed power-law and APEC
models (\S\ref{s:bright}).  Note that the \chandra and \xmm spectra
were taken much earlier than the \nustar observations. The
best-fit normalizations relative to \nustar are listed in
Table~\ref{t:src:fit}.

We also explored the \chandra spectra of the 15 \nustar sources with $>$200
net counts in the \chandra 0.3--8 keV band to constrain the presence of
the iron lines at 6.4 and 6.7 keVs.  We fit the \chandra spectra with and
without the iron lines for an absorbed power-law model (\S\ref{s:mcv}). 
The sources showing the iron lines are flagged with ``{\it f}"
in Table~\ref{t:ap}.

\subsection{X-ray Variability} \label{s:var}

The millisecond time resolution of the \nustar FPMs allows us to characterize
the timing properties of detected sources over a range of timescales. The
\nustar timing resolution is $\sim 2$~ms rms, after corrected for
thermal drift of the on-board clock, and the absolute accuracy is known
to be better than 3~ms \citep{Mori14, Madsen15}.  In our search for
periodic modulations (see below), all photon arrival times are converted to
barycentric dynamical time (TDB) using the \chandra coordinates of each
point source.

To characterize the source variability we used the KS statistic to
compare the temporal distributions of X-ray events extracted from
source and background apertures in the 3--40~keV energy band of each
observation. The background lightcurve acts as a model for the expected
source counts as a function of time. The maximal difference between
two cumulative normalized lightcurves gives the probability that they
are drawn from the same distribution. i.e., that the source tracks the
background.  If the probabilities of the KS statistics are less
than $\sim$\nep{3.8}{-5}, which is equivalent to 1\% random chance
probability after taking into account the number of search trials
(i.e.~the sum of the number of observations searched for each source),
we consider the source as variable and it is flagged with
"{\it k}" in Table~\ref{t:ap}. We manually checked
the source and background lightcurve for candidate variable sources
to avoid the false detection due to background fluctuation.
NGP 2 is the only source showing 
significant variablility.

In the case of variability from observation to observation, in order to
account for the large differences in the off-axis responses
among multiple observations of a given source, we compare the observed
fluxes of each source calculated under the proper response function of
each observation and use the maximum-to-minimum flux ratio ($r$) as an
indicator of the variability.
Table~\ref{t:src:var} lists sources with multiple observations that show
possible flux variability.  A caveat is that the error of the flux ratio
is in general dominated by the relatively large uncertainty of the minimum
flux value, which often implies $r$ being statistically consistent with 1
(i.e.~no variability) even for the cases with $r$ $\gg$ 1 (e.g.~NGP 66).
Columns 7 and 8 in Table~\ref{t:src:var} show the lower limit of
the observed flux ratio equivalent to 1 and 2$\sigma$, respectively.
Many of these limits are very close to 1 even
though these limits do not account for the number of the search trials
(67 sources with multiple observations).

In order to evaluate the significance of the observed flux ratios,
we calculate two random chance probabilities for each soure under the
assumption of the source flux being constant: a probability for observing
a higher-than maximum flux and a lower-than minimum flux (column 9) and
a probability for having the flux ratio greater than the observed flux
ratio (column 10).  The probabilities in the table are without accounting
for the search trial numbers.  The former is more binding and thus less
probable than the latter since the former uses specific flux values in
calculating the probability, and as a result, it is much more sensitive
to the accuracy of the mean flux estimate than the latter.  The total
number of the search trials in the two are also different: in the former
it is proportional to the total number of the searched observations
(e.g.~10\sS{-4.4} in column 9 is equivalent to a true random probability
of $\sim$1\% after accounting for the trial numbers), whereas in the
latter it is proportional to the number of the searched sources
(e.g.~10\sS{-3.8} in column 10 is equivalent a true random probability
of $\sim$1\%). 

Table~\ref{t:src:var} shows that it is not unusual to observe a high flux
ratio ($\gg$1) even for a constant flux source, depending on the photon
statistics. The three sources NGP 2 (\S\ref{s:GRS1741}), 4
(\S\ref{s:6369}) and 7 (\S\ref{s:5908}) show very significant flux
variations under both scenarios of the random chance probabilities, and they
are flagged with {\it r}.  The observed flux ratios of the other sources
appear statistically probable even if their X-ray emission is actually
steady, but the large deviation of observed minimum and maximum fluxes
relative to the mean values may imply some degree of the flux variation.

We also searched for a pulsar signal from those \nustar sources with
sufficient counts to detect a coherent timing signal, determined
as follows.  The ability to detect pulsations depends strongly on
the source and background counts and number of search trials.  For a
sinusoidal signal, the aperture counts (source plus background) necessary
to detect a signal of pulsed fraction $f_p$ is $N=2S/f^2_p$, where $S$
is power associated with the single trial false
detection probability of a test signal $\wp = e^{-S/2}$ and is
distributed as $\chi^2$ with two degrees of freedom.  In practice, for
a blind search, we need to take into account the number of frequencies
tested $N$\Ss{trials} = $T$\Ss{span} $f$\Ss{Nyq}, when $T$\Ss{span} is the data
span and $f$\Ss{Nyq} = 250~Hz, the effective \nustar Nyquist
frequency. In computing $f_p$ we must allow for the reduced
sensitivity of the search due to background contamination in the
source aperture ($N_b$); the minimum detectable pulse fraction
$f_p$\Ss{,min} is then increased by $(N_s+N_b)/N_s$ where $N_s$ is the source
counts.

We computed the pulsar signal detectability in individual observations
for each source in our sample and find that
nearly all sources proved undetectable even if their flux were 100\% pulsed.
However, we have identified four sources for which we can potentially 
place a meaningful
limit of $f_p$\Ss{,min} $<$ 50\% on the pulsed flux, at the 3$\sigma$
confidence level.  These are the first four entries in the bright source
list below, NGP~1--4.  For each source we evaluated the power at each
frequency (oversampling by a factor of two) using the unbinned $Z^2_n$
test statistic \citep{Buccheri83} summed over
$n$ = 1, 2, 3 and 5 harmonics, to be sensitive to both broad and narrow pulse
profiles.  We initially searched photon arrival times with energies in
the 3--40 keV range and used an nominal $30^{\prime\prime}$
aperture. We repeated our search for an additional combination of
energy ranges 3--25~keV, 3--10~keV, 10--25~keV, and
10--40~keV, and for aperture sizes of $r< 20^{\prime\prime}$ and $r<
30^{\prime\prime}$. For all these searches no significant signal was
detected. We found $f_p$~$<$~6.1\% for NGP 1 from one observation 
and $<$~8.0\% for NGP 2 (the best out of the three observations)
at 3$\sigma$, and the other two sources (NGPs 3 and 4), where
the search was divided into three observations for each source,
did not produce a meaningful upper limit on the pulsed fraction.

\section{Bright X-ray Sources} \label{s:bright}

In this section, we review the broadband X-ray properties of nine bright
\nustar sources, including four sources for which detailed analyses
of the \nustar observations are found in the literature.  We analyze
the broadband spectra of four other \nustar sources using the \chandra
and \xmm archival data, and comment on another bright X-ray source detected
by \nustar.

\subsection{NGP 1 (1E 1743.1-2843 or \#7722)}\label{s:1E1743}

1E1743.1-2843 was discovered by the {\it Einstein Observatory} more
than three decades ago \citep{Watson81} but the precise nature of the source
remains unclear.  \citet{Lotti15} present the results of 
recent \nustar and \xmm observations of the source. They concluded that
between two proposed scenarios, LMXB or HMXB,
it is likely a LMXB based on the argument that the absence of periodic
pulsations, eclipses or the Fe K\Ss{\alpha} line in the X-ray emission
disfavors the HMXB scenario more strongly.
X-ray spectral model fitting requires a composite model, which includes
a disk blackbody and a cut-off power-law component.
For an absorbed power-law model, the median energy is consistent
with $\Gamma$ = 1.9 $\pm$ 0.2 and the quantile analysis (see \S\ref{s:class} and
Fig.~\ref{f:qd}) favors a thermal plasma model indicating a strong
thermal component in the X-ray emission. 
It was the brightest source in our survey of the GC region with an
absorbed photon flux reaching $\sim$2$\times$10\sS{-2} \pcgs in the
3--40 keV band. 
\citet{Lotti15} estimate a luminosity of
$L$\Ss{2-10\ keV}$\sim$10\sS{36} \lcgs at 8 kpc, which is consistent with
our estimate: $L$\Ss{3-10\ keV}$\sim$1.3$\pm$0.2$\times$10\sS{36}
\lcgs within 2$\sigma$.

\subsection{NGP 2 (GRS 1741.9--2853 or \#6090)} \label{s:GRS1741}

Since its discovery by the {\it Granat} satellite \citep{Sunyaev90}, the
transient X-ray source GRS 1741.9--2853 (AX~J1745.0--2855), has
produced at least a dozen Type I outbursts, typical of LMXBs
binaries \citep{Cocchi99}, recorded by several X-ray telescopes over
the years \citep[see][and references
therein for a review]{Degenaar14}. \nustar observed GRS 1741.9--2853
four times, during one of which a Type I
burst was fully recorded.  A comprehensive paper on these data sets
is presented in \citet{Barriere15}. These authors were able to place
a lower limit of 6.3~kpc on the distance to the NS based on the peak
flux from the burst assuming the photospheric radius expansion model.
They argue that spectral variation during outburst suggests disturbances
in the inner accretion disk resulting from the burst. In the work herein
we exclude a 352~s burst interval and report our analysis results in
Table~\ref{t:ap}.
Table~\ref{t:src:var} shows a significance flux variation by nearly
three orders of magnitude and during a quiescent period the flux fell
below the detection level.

We generally reproduce the earlier result. The median energy of the
X-ray spectrum is consistent with an absorbed power-law model with
$\Gamma$ = 2.6 $\pm$ 0.3. The quantile diagram indicates that
the overall X-ray spectrum, which is still dominated by the outbursts
even after the exclusion of the peak burst period, is more consistent with
blackbody emission than a power-law model (Fig.~\ref{f:qd} in
\S\ref{s:class}), as expected, since the thermal emission from the
surface becomes dominant during the outburst periods.  

For a timing analysis we considered the quiescent, outburst, and burst
intervals separately.  The source and background counts combination for
each interval allows for a well constrained pulsar search. A comprehensive
search did not produce a significant signal for any interval, consistent
with the null timing search result reported in
\citet{Barriere15}.

\begin{figure*} \begin{center}
\includegraphics*[width=0.99\textwidth]{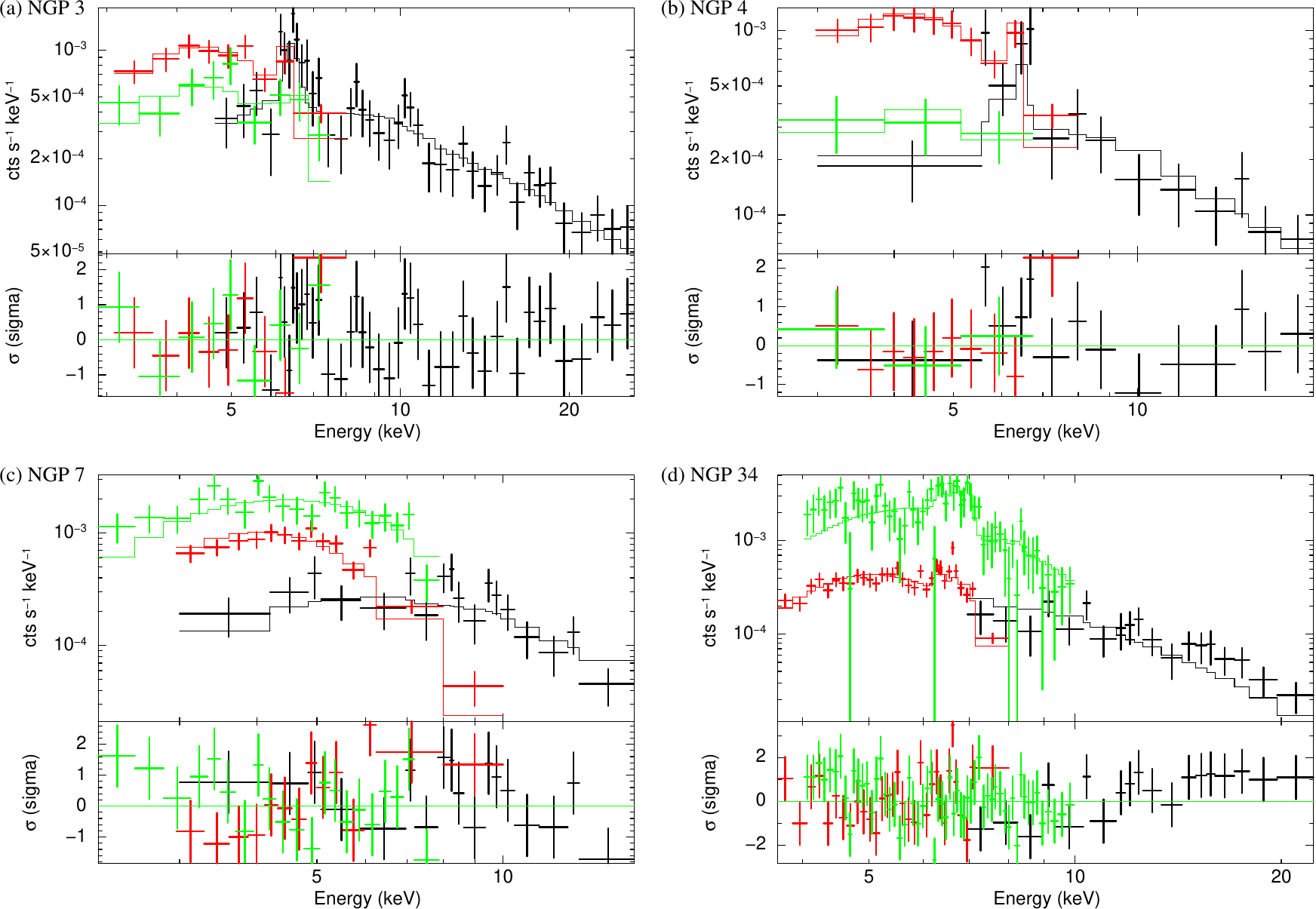}
\caption{Joint spectral model fits of four bright \nustar sources: 
(a) NGP 3 or \chandra ID \#4942 (see \S\ref{s:4942}),
(b) NGP 4 or \#6369 (\S\ref{s:6369}),
(c) NGP 7 or \#5908 (\S\ref{s:5908}),
(d) NGP 34 or \#1568 (\S\ref{s:1568}).
The \chandra, \xmm and \nustar spectra are shown in red, green and black, respectively.
The solid lines are the best-fit model for an absorbed power-law model  in (a), (b) and (c),
and for an absorbed APEC model in (d). The models for (a), (b) and (d) include a 6.4 keV 
neutral Fe line. See Table~\ref{t:src:fit} for the best-fit parameters.
}
\label{f:spec}
\end{center}
\end{figure*}

\subsection{NGP 3 (CXOUGC J174413.7-285423 or \#4942)} \label{s:4942}

NGP 3 is a bright, very hard X-ray source in block A with a median energy
of $\sim$ 10 keV. It is one of a few sources that are detected above 20
keV and the 3rd brightest sources in the 10--40 keV band with an X-ray
luminosity of \nep{1.4}{34} \lcgs.  The measured X-ray fluxes of the
source appear to vary by 60\% between two observations, about 14
months apart, but such a variation in the measurement is statistically
plausible even for a constant flux source (i.e.~70\% chance to see
such a variation from a source when accounting for the search trials,
Table~\ref{t:src:var}).
Fig.~\ref{f:spec}a shows a joint
model fit of non-simultaneous \nustar (black), \chandra (red) and
\xmm (green) spectra for an absorbed power-law model with a
Gaussian line fixed at 6.4 keV. The best-fit photon index is 0.9 $\pm$ 0.3 and
the Gaussian line
improves the fit (from $\chi_r^2$ = 1.49 to 0.96) with the best-fit
equivalent width (EW) of 770 eV (Table~\ref{t:src:fit}). 
For single temperature thermal models, 
the plasma temperature is not well constrained but the best-fit plasma temperature
for an absorbed APEC model 
is $\gtrsim$ 30 keV with 95\% confidence.
The 3--40 keV absorbed luminosity at 8 kpc estimated by the spectral fit is
\nep{1.1-1.6}{34} \lcgs, which is consistent with the
aperture photometry result, \nep{1.6}{34} \lcgs.
Given the hard continuum in the X-ray spectrum with the neutral iron line,
we suspect that NGP 3 is most likely an IP, although the observed 
X-ray luminosity at 8 kpc is at the high end of the luminosity
distribution for IPs.

\subsection{NGP 4 (CXOUGC J174515.6-284512 or \#6369)} \label{s:6369}

NGP 4 is another bright X-ray source in block A with a median energy
of 8.2 keV. It is also detected in the 20--40 keV band. Four
observations covered the source and we excluded one of the module B
data due to the SL background. The X-ray flux of NGP 4
varied by about a factor of two over about a year, which has 
$\sim$ 3\% random chance probability when accounting for
the 67 sources searched for the variability (Table~\ref{t:src:var}).
Fig.~\ref{f:spec}b shows a joint model fit of the \nustar (black),
\chandra (red) and \xmm (green) spectra for an absorbed power-law model
with a Gaussian line fixed at 6.4 keV.  The best-fit photon
index and EW are 1.1 $\pm$ 0.5 and 620 eV, respectively.  The spectral model fit
requires the Gaussian line, otherwise the reduced $\chi^2_r$ increases to 1.5.
For thermal plasma models, the spectral fit does not constrain
the plasma temperature.  The 3--40 keV absorbed luminosity estimated
by the spectral fit is \nep{6.1-8.3}{33} \lcgs, which is consistent with
the aperture photometry estimate, \nep{8.1}{33} \lcgs.
This source is also suspected to be an IP.

\subsection{NGP 7 (CXOUGC J174454.1-285842 or \#5908)} \label{s:5908}

NGP 7 is located in the overlapping section of the mini survey, blocks A and B.
As a result, seven observations covered the source, but we excluded two
observations and the FPM B of another, which did not contribute much.
The X-ray flux shows the 2nd largest variation after 
NGP 2, changing by a factor of four over two years.   Constant, steady X-ray
emission from the source is statistically ruled out (Table \ref{t:src:var}).
Fig.~\ref{f:spec}c shows a joint model fit of non-simultaneous \nustar (black),
\chandra (red) and \xmm (green) spectra for an absorbed power-law model.
The spectral fit does not require any iron lines. The best-fit photon
index is 1.2 $\pm$ 0.4. In the case of an absorbed APEC model, the plasma
temperature is poorly constrained but the best-fit temperature is significantly
lower than NGPs 3 and 4.  The 3--40 keV luminosity at 8 kpc estimated by
the spectral fit and aperture
photometry are \nep{6.0-9.6}{33} and \nep{6.2}{33} \lcgs, respectively.
Given the photon index, the lack of the neutral iron
line in the X-ray spectrum and the large X-ray variability, we suspect
the source is a quiescent NS or BH X-ray binary or a background
active galactic nucleus (AGN).

\subsection{NGP 8 (The Cannonball or \#2743)} \label{s:cannonball}

The Cannonball, discovered by \chandra in 2003 \citep{Muno03},  
is likely a run-away pulsar 2\arcmin\ northeast of Sgr~A*,
just outside the radio shell of the supernova remnant \citep{Zhao13}.
The cometary emission surrounding the source is interpreted as a pulsar
wind nebula (PWN) and the projected velocity
is estimated about 500 km s\sS{-1} \citep{Park05, Zhao13}, but no pulsation has been
detected so far.  The detailed spectral analysis
of the \nustar observation of the source can be found in \citet{Nynka13}.
They observed a non-thermal component up to 30 keV in the X-ray
spectrum, which is described by an absorbed power-law model
with $\Gamma$ = 1.6 $\pm$ 0.4 and 
\nH = 3.2$\times$10\sS{23} cm\sS{-2}. Their estimate of the absorption is 
about 5 times larger than the typical interstellar absorption
assumed in the GC region, and also higher than the estimates based on
$A_{Ks}$ (\S\ref{s:class}). The high
extinction is consistent with the idea of the local absorption caused
by the surrounding PWN. 
Our estimate of the photon index ($\Gamma$ $\sim$1.8 $\pm$ 0.2) under the assumption of
\nH = 6$\times$10\sS{22} cm\sS{-2} still matches theirs within the uncertainty,
reconfirming the presence of the non-thermal emission above 10 keV.
The unabsorbed X-ray luminosity in
the 2--30 keV band is about 1.3$\times$10\sS{34} \lcgs according
to \citet{Nynka13}.  The corresponding observed luminosity in 3--40
keV is about 10\sS{34} \lcgs,
which is consistent with our estimate, $\sim$9$\times$10\sS{33} \lcgs.

Given the complex diffuse emission surrounding the source, 
the background aperture has to be carefully selected
as discussed in \S\ref{s:aperture}. Despite the significant difference
in aperture selection between our analysis and \citet{Nynka13}, the
consistent results between the two are encouraging and indirectly validate
our aperture photometry procedure. 

\subsection{NGP 20 (GRO J1744--28 or \#5436)} \label{s:gro1744}

GRO J1744--28 was discovered in 1996 as a transient source by
the Burst and Transient Source Experiment (BATSE) on
board the {\it Compton Gamma-Ray Observatory} \citep{Kouveliotou96}.
It is an LMXB with multiple Type~II X-ray bursts and
named as the Bursting Pulsar since it exhibits both bursts and pulsations 
(2.14 Hz with the orbital period of 11.8 days). \citet{Younes15} present the
analysis results of a simultaneous \chandra and \nustar observation
during an outburst on 2014 March 3, which was the 3rd occurrence since
its discovery. They detected the X-ray emission up to 60 keV at the
Eddington flux level or higher, and the spectrum is well described by
a blackbody plus a power-law model with an exponential cut-off.

In our survey the source was observed in 2013 August and July and again
in 2014 August when it was relatively quiescent with no significant X-ray
emission above 10 keV. The 3--10 keV X-ray luminosity at 8 kpc was about
2$\times$10\sS{33} \lcgs. According to quantile analysis
the spectrum was consistent with $\Gamma$ = 3.1 $\pm$ 0.5 for an
absorbed power-law model, making it the softest source among the \nustar
detections in the GC region.  This result is consistent with 
the earlier \chandra \citep{Wijnands02} and \xmm \citep{Daigne02}
observations of the source in quiescent states,
where they found the 0.5--10 keV X-ray luminosity of
$\sim$3$\times$10\sS{33} \lcgs and the photon index of $\Gamma$\Ss{S} = 2--5.

\subsection{NGP 34 (CXOUGC J174534.5-290201 or \#1568)} \label{s:1568}

NGP 34 is one of the two bright hard X-ray sources (the other is NGP
31) found just on the western edge of the Sgr~A diffuse complex.  NGP 34 was
covered by seven observations, and three observations collected more than 100
net counts for the source.  A main challenge for NGP 34 is
in handling the diffuse background where the selection of the
background aperture becomes critical (the bottom panel in
Fig.~\ref{f:example}). For stacking the individual spectra of the
multiple observations, we use the same aperture regions used for aperture photometry. 

Fig.~\ref{f:spec}d shows a joint model fit of non-simultaneous \nustar (black),
\chandra (red) and \xmm (green) spectra for an absorbed APEC model with
a Gaussian line at 6.4 keV. The reduced $\chi^2_r$ is about 1.2. Including a
partial covering component, which is commonly used for describing X-ray spectra
from IPs \citep{Hailey16}, improves the fit 
($\chi^2_r$$\sim$1.1), but the parameters for the partial covering
component are not well constrained.
The plasma temperature is found to be 8.6 and 12 keV with and without partial
covering, respectively.  The X-ray spectrum in the 6--7 keV band shows a clear sign
of additional line emission besides the neutral iron line, which is also
consistent with the lower plasma temperature than NGPs 3 and 4.

An absorbed power-law model fits the spectra relatively
poorly even with an iron line at 6.4 keV ($\chi^2$$\sim$1.5, $\Gamma$ $\sim$ 
1.5).  The median energy of 11 keV (Fig.~\ref{f:qd}) translates 
to $\Gamma$ = 0.7 $\pm$ 0.2 for an absorbed power-law model, but the high
quartile ratio also implies that a thermal plasma model may be better
suited for the source.  Both spectral model fit and quantile analysis
estimate similar 3--40 keV photon fluxes of \nep{2.2}{-5} and
\nep{2.6}{-5} \pcgs, respectively.  The 3--10 keV absorbed
luminosity is \nep{0.6-1.1}{33} \lcgs at 8 kpc.
The hard X-ray spectrum with the strong iron lines
suggests that NGP 34 is likely an IP.


\begin{figure*} \begin{center}
\includegraphics*[width=0.99\textwidth]{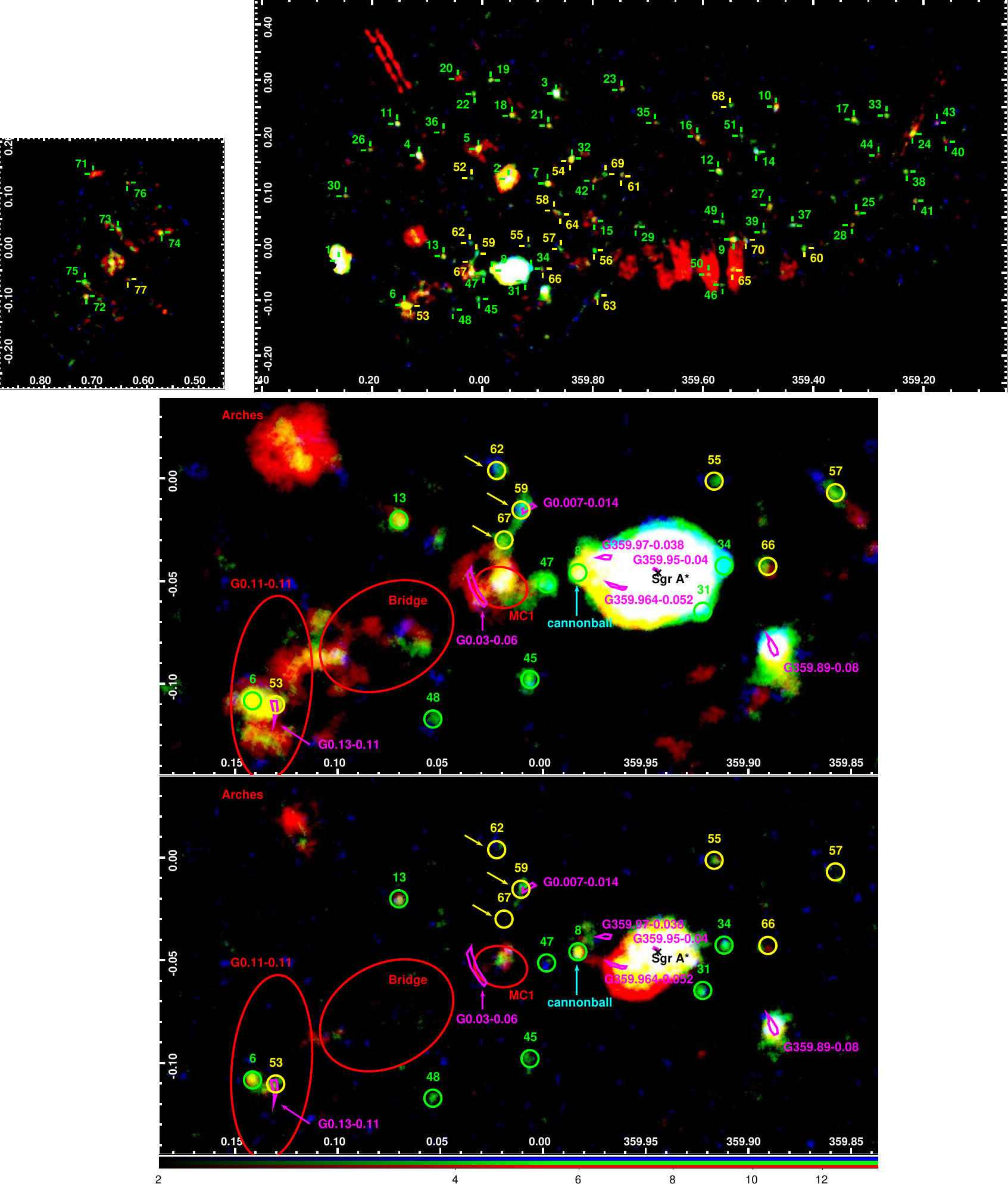}
\caption{Three-color trial maps of the GC region using 30\% PSF enclosures
(top), the region around Sgr~A* using 30\% (middle) and 15\% (bottom)
enclosures.  The color setting: red=3--10 keV, green=10--20 keV and blue=20--40 keV.
The group-1 and 2 sources are marked in green and yellow,
respectively and they are labeled with the \nustar source IDs.  Some of
the molecular clouds and the X-ray filaments are marked with red
ellipses and magenta polygons, respectively \citep[see][]{Ponti15a}. The
yellow arrows point the three hard X-ray sources without clear soft
X-ray counterparts (\S\ref{s:hardsources}).
}
\label{f:sgra}
\end{center}
\end{figure*}

\subsection{NGP 46 (KS 1741--293 or \#5835)} \label{s:5835}

KS 1741--293, discovered in 1989 by the X-ray wide field camera
TTM on the Kvant module of the Mir space station \citep{intZand91}, is
a transient NS LMXB, exhibiting Type I bursts.
In the hard X-ray band above  20 keV,  the source
was detected by \integral for the first time \citep[third IBIS
catalogue by][]{Bird07}. \citet{Marti07} misidentified CXOUGC
J174451.0-292116 (\#5824) as the \chandra counterpart of the source 
because at the time it was only the \chandra source 
consistent with the positions of the previous detections. 
The subsequent transient activities from KS 1741--293
\citep{Degenaar13} indicate that the real \chandra counterpart is
CXOUGC 174451.6-292042 (\#5835), which is located about 4\arcsec\ from
NGP 46.
\citet{Degenaar13} show the \chandra and \swift 2--10 keV flux 
of the source varies from \nep{6}{-14} to \nep{2}{-10} \fcgs while
the photon index varies from $<$ 1 to $>$ 2.
\citet{DeCesare07} reported a 2 yr monitoring of the source with
\integral from 2003 February to 2005 May, where they observed that the hard X-ray
emission above 20 keV from the source also varied by more than a factor
of 10 and reached as high as 20 mCrab ($\gtrsim$10\sS{36} \lcgs at 8
kpc) in the 15--30 keV band. The 5--100 keV broadband JEM-X and
IBIS/ISGRI spectra were well fitted by a disk blackbody plus a cut-off
power-law or a Comptonized model.  In our survey, the source was
covered by a single 50 ks observation in 2013 September. 
Unfortunately the source fell between two GR streaks of the
bright neighbor 1A 1742--294.  Thus its soft X-ray emission below 10 keV is
somewhat uncertain, but its quiescent hard X-ray emission above 10 keV
was well detected with $L_X$~$\sim$~\nep{8.9}{33} \lcgs at 8 kpc.

\section{Discussion}

We have discovered 77 hard X-ray sources from the first \nustar survey of the
GC region.  For source detection, we introduced trial maps - new
detection significance maps based on Poisson statistics-driven random
chance probabilities. In \S\ref{s:hardsources} we explore unusually hard X-ray sources
found in the trial maps of the GC region. In
\S\ref{s:sensitivity} we estimate the overall survey
sensitivity.  In \S\ref{s:missing} we study the significance in the lack of
foreground sources in our
survey.  In \S\ref{s:lnls} we calculate the \lnls distributions of the
\nustar sources and illustrate how these \nustar
results break some of the spectral degeneracy seen in the
\chandra observations. 
In \S\S\ref{s:mcv} and \ref{s:msp}, we explore the nature of the hard
X-ray sources in the GC region with two possible source types - MCVs
and rotationally powered pulsars.

\subsection{Unusually Hard X-ray Sources in the GC region}
\label{s:hardsources}

Fig.~\ref{f:sgra} shows three-color trial maps (red: 3--10 keV, green:
10--20 keV, blue: 20--40 keV) of the GC region and a close-up region
around Sgr~A*.  The bright X-ray emission from many
diffuse and point sources
near Sgr~A* saturates the image in its immediate neighborhood.
The trial map around Sgr~A*  revealed a cluster of hard X-ray sources
(NGPs 59, 62 and 67; the yellow arrows in the bottom two panels of
Fig.~\ref{f:sgra}) in the north of a Sgr~A molecular cloud, MC1.
These hard
X-ray sources do not have obvious soft X-ray counterparts,
and thus the nearby brightest and closest \chandra sources 
(CXOUGC J174542.3--285606, J174539.5--285453 and J174546.9--285608)
are assigned to be potential counterparts.
According to the quantile diagram in Fig.~\ref{f:qd}, these sources are
unusually hard with $\Gamma$ $<$ 1 for a power-law model or $kT$$>$
50 keV for a single temperature thermal plasma model.  

In particular, NGP 59 is located at the southern end of the small
(11\arcsec$\times$6.5\arcsec) X-ray filament,
G0.007--0.014 \citep{Johnson09, Ponti15b}. According to
\citet{Johnson09}, the soft ($<$ 10 keV) X-ray spectrum of the
filament has a photon index
of $\Gamma$\Ss{S} $\sim$ 1 for a power-law model (albeit with a large uncertainty) and the
2--10 keV luminosity is $\sim$2$\times$10\sS{32} \lcgs.  This is
consistent with our aperture photometry results of the \nustar source
(i.e.~no detection below 10 keV).
Therefore, we cannot rule out the X-ray filament as the origin of the
observed hard X-ray emission.  
The broadband (3--40 keV) spectrum of NGP 59
shows $\Gamma$ = 0.0 $\pm$ 0.2 for an absorbed power-law model.
For comparison, G359.97--0.038 and Sgr~A--E, two prominent non-thermal
filaments in the region, show $\Gamma$ = 1.3 and 2.3,
respectively \citep{Nynka15, Zhang14}.
If the \nustar detection is indeed from
the X-ray filament, this is the first detection of its kind with 
such dominant hard ($>$10 keV) X-ray emission.

Besides these three sources, about a dozen \nustar
sources exhibit extremely hard X-ray spectra
(i.e.~median energies $\gtrsim$ 9 keV or $\Gamma$ $<$ 1 in
Fig.~\ref{f:qd}, or blue sources in Fig.~\ref{f:sgra}).  
Some of these are suspected to be IPs with relatively high plasma
temperatures \citep[e.g.~NGP 3, see \S\ref{s:4942} and \S\ref{s:lnls};
see also][]{Perez15,Hailey16}.
Of the 77 \nustar sources,
66 sources show significant X-ray emission in hard ($>$ 10 keV) X-ray bands
(column 8 in Table~\ref{t:src}).

In the hard X-ray band above 40 keV, only two significant objects,
both near Sgr~A*,
are observed. \citet{Mori15} explored these in
the 40--79 keV trial maps generated from three observations of the Sgr~A*
field. One of the objects coincides with the head of G359.95--0.04,
a PWN, and the other, detected at $\sim$4$\sigma$ and a bit elongated in shape, 
does not seem to have
a clear counterpart in the \chandra and \xmm images.  The trial maps
of the full survey data show a similar result, but the morphology of
the 2nd source appears less elongated.

\subsection{Survey Sensitivity} \label{s:sensitivity}

We follow the recipe by \citet{Georgakakis08} to estimate the sensitivity
limit and the sky coverage of the survey. They calculated Poisson
statistics-based cumulative detection probabilities expressed in an
incomplete Gamma function, which is basically the same formula as Eq.~1.
Thus, their approach is appropriate for our source search method.

\begin{figure} \begin{center}
\includegraphics*[width=0.45\textwidth,clip=true] {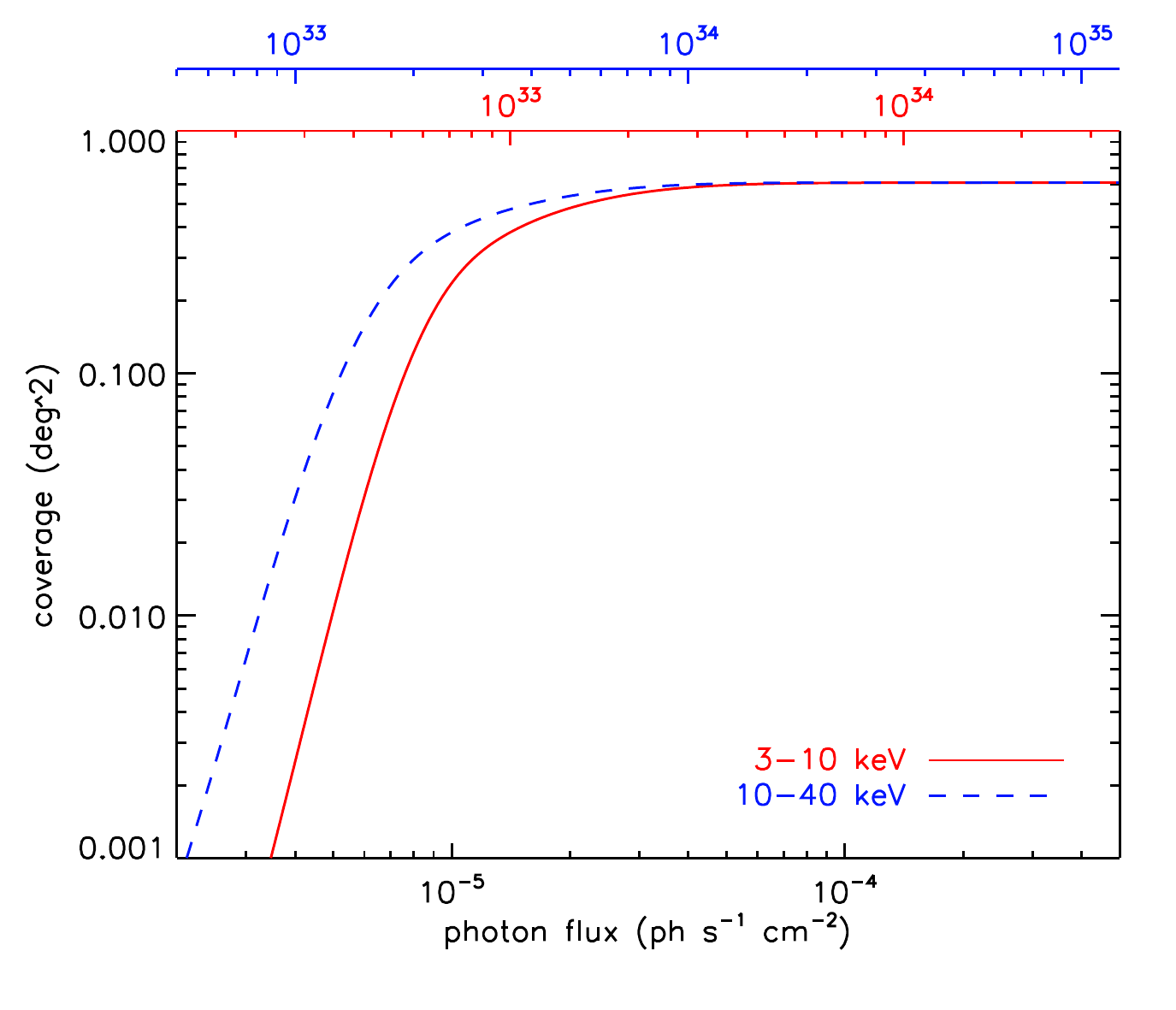}
\caption{Sky coverage of the main GC region
in the 3--10 and 10--40 keV bands as a function of absorbed photon flux.
The top $x$-axis shows the corresponding X-ray luminosities at 8 kpc
for each band, based on the source-averaged conversion factor
(\S\ref{s:lnls}).
}
\label{f:skyc}
\end{center}
\end{figure}

\begin{figure*} \begin{center}
\includegraphics*[width=0.99\textwidth,clip=true] {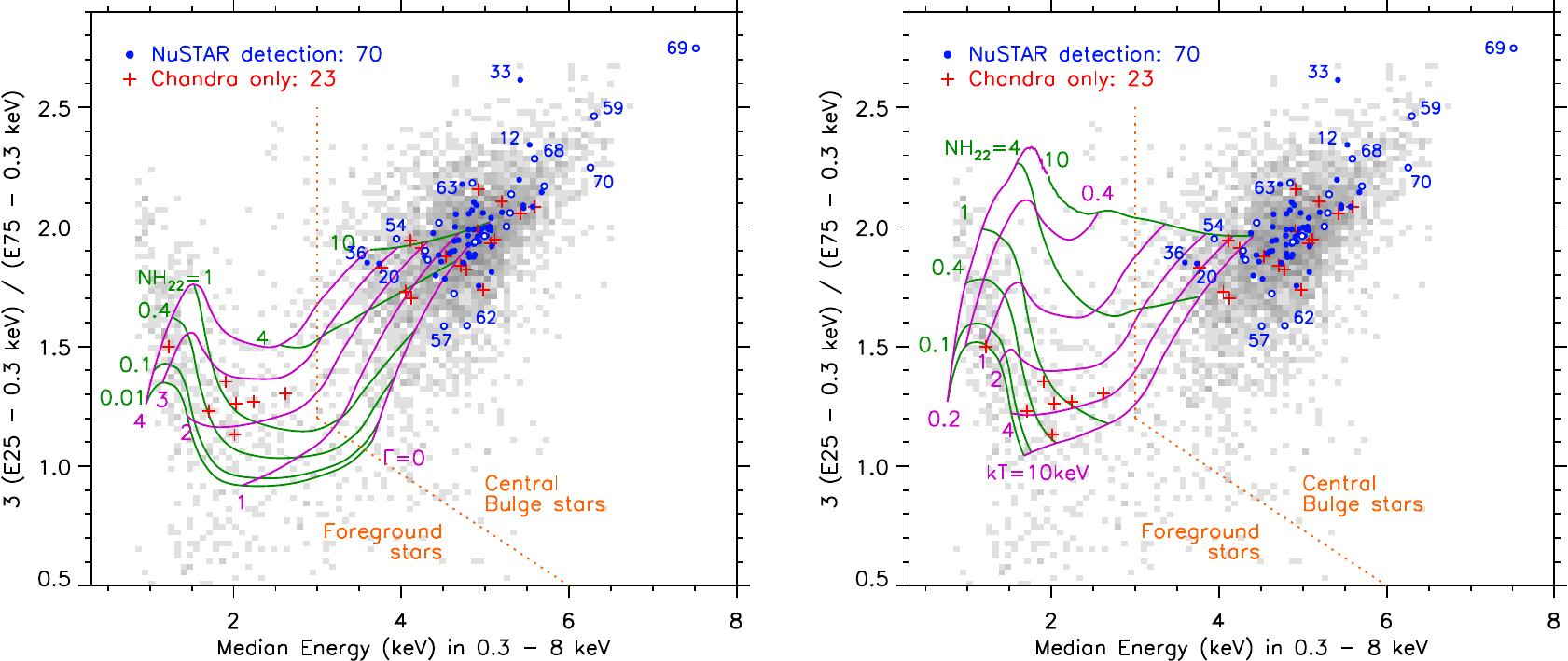}
\caption{\chandra 0.3--8 keV quantile diagrams of the
\chandra sources in the \nustar survey region.
The grey density map shows the relative \chandra source
distribution.
The \nustar detections are marked with the 
(blue) closed circles for group 1 and the (blue) open circles for
group 2.  The (red) crosses show the \chandra sources with no \nustar
detections but they are in relatively uncrowded regions and their
\chandra 2--8 keV fluxes are high enough for \nustar detections.
The grids are for power-law models with 
photon index $\Gamma$\Ss{S} = 0, 1, 2, 3, and 4 (left) and 
for thermal bremsstrahlung models with $kT$ = 0.2, 0.4, 1, 2, 4,  and 10 keV (right). 
The grids also cover \nH = 0.01, 0.1, 0.4, 1, 4, and 10\x10\sS{22} cm\sS{-2}.
The (orange) dotted lines roughly separate the foreground sources
from the central bulge and background AGN sources.
}
\label{f:qd_chandra}
\end{center}
\end{figure*}

For a given detection threshold ($P_T$), we first find the matching threshold for the total
counts ($N_T$), then we can calculate the detection probability
that a source with a given flux ($f$) generates the counts more than
both $N_T$ and the observed counts ($N^*$).  

\begin{eqnarray}
 	& P(N>N_T | &\lambda_S=0, \lambda_B)  =   P_T, \\
 	& P(N>N_0 | &\lambda_{S_f}, \lambda_B)  = 
		\frac{1}{\Gamma(N_0+1)} \int^{\lambda_B+\lambda_{S_f}}_0 e^{-t} t^{N_0} dt \label{e:sensitivity}
\end{eqnarray}
where $N_0$ is max($N_T$, $N^*$) and $\lambda_{S_f}$ is the 
mean counts expected from a source with flux $f$.
For a given flux $f$, we calculate the probability $P(N>N_0|\lambda_{S_f},
\lambda_B)$ for every pixel, and the sky coverage is given by the
probability sum over all the pixels.\footnote{A small difference in
the normalization [$\Gamma(N+1)$ vs.~$\Gamma(N)$]
between the above formula and \citet{Georgakakis08} is from the fact
that we use $P(N>N^*)$ for both source detection and sensitivity
calculation whereas \citet{Georgakakis08} use $P(N\ge N^*)$. As long
as a consistent
normalization is used for both source detection and sensitivity calculation,
either normalization is valid.}
We repeat the calculation as a function of photon flux. 

The observed (or absorbed) photon flux ($f$) is calculated as
as $f=\lambda_{S_f}/ (T A)$ where $T$ is the exposure time and $A$
the effective area.  The exposure time of each sky pixel is given in the
vignetting-free exposure mosaic.  For the effective area, we generate
an exposure map vignetted for each energy in 1 keV steps. Then for
each pixel, we sum up the effective area of each energy over a given
energy band, weighted by the stacked energy histogram of all the sources.
In this way, for every pixel in a given energy band, we can calculate
the source-spectrum averaged conversion factor from photon counts to flux.

Since we use three detection cell sizes, we use the largest detection
probabilities of the three cases to get a collective sky coverage in each
band. Fig.~\ref{f:skyc} shows the resulting sky coverages as a
function of photon flux in the 3--10 and 10--40 keV bands. 
The top $x$-axis shows the matching X-ray luminosities at 8 kpc
using the source-averaged flux to luminosity conversion factor
(\S\ref{s:lnls}).
The survey covers about 0.01 deg\sS{2} at
$\sim$ 3--4\x 10\sS{32} \lcgs and 0.6 deg\sS{2} above $\sim$ 2\x10\sS{33} \lcgs
in the 3--10 keV band.
In 10--40 keV, it covers about 0.01 deg\sS{2} at 
$\sim$ 8--9\x 10\sS{32} \lcgs and 0.6 deg\sS{2} above $\sim$ 5\x10\sS{33} \lcgs.

\subsection{Missing Foreground Sources?} \label{s:missing}

Fig.~\ref{f:qd_chandra} shows the \chandra quantile diagrams in 0.3--8 keV with
power-law (left) and thermal bremsstrahlung (right) model grids. 
The (grey) density map
indicates the general distribution of \chandra sources with
$\gtrsim$ 50 net counts in the \nustar survey region.  
A large cluster of the sources around a median energy of 5
keV and a quartile ratio of 2, where \nH $\gtrsim$ \nep{4}{22}
cm\sS{-2}, are either in the central Galactic Bulge
or background AGN.
The (blue) circles show the \chandra counterparts of the \nustar detections:
the closed and open circles are for the group-1 and 
2 sources, respectively.  The (red) crosses indicate about two dozen \chandra
sources without \nustar detections, but their  \chandra 2--8 keV flux
should have been high enough for the \nustar detections, and they are located 
in relatively confusion-free sections of the survey region.
Missing these
relatively bright sources in the \nustar survey is not particularly
surprising given the X-ray flux variability of the \chandra
sources, but the relative ratio between the foreground and the central
bulge sources is intriguing.

The diagram indicates that all the \nustar detections are either in the central
bulge near the GC or background AGN.  This appears true even for the
group-2 sources whose \chandra counterparts are a
bit tentative.  It implies that many of them are indeed true counterparts.
The lack of the foreground sources in the \nustar detections
 contrasts with the fact that 30\% of all the \chandra
sources in the region (the grey density map) or 30\% of 22 relatively
bright \chandra sources without \nustar detections (red crosses) are
foreground sources.  It shows that the \nustar selected X-ray sources
in the GC region have an intrinsically harder spectral distribution than the
foreground X-ray source population detected by \chandra in the region.

The \nustar sources provide a unique, unobscured view of the Galactic
X-ray source
population from the local solar neighborhood to the central bulge 
since the interstellar absorption to the GC (10\sS{22}--10\sS{23}
cm\sS{-2}) has little effect in the \nustar bands. 
One can test if
the \nustar selected X-ray sources follow the stellar
population by calculating how many foreground sources we should
have detected for a given detection of a central bulge source.
We consider sources within 4 kpc of the Sun
(i.e.~\nH$\lesssim$3$\times$10\sS{22} cm\sS{-2}) as foreground stars
and sources at distances of 6--10 kpc as bulge stars since
the interstellar absorption peaks at around 4--5 kpc toward the GC
\citep{Drimmel03}.
Then the relative ratio of the stellar volume density between
the foreground and central bulge sources within the survey FoV is about 0.034\%
according to the stellar distribution model used in \citet{Muno06} and
\citet{Hong09}.
Assuming that the foreground sources are closer than
the central bulge sources by a factor of four on average, 
if the cumulative X-ray luminosity distribution follows a slope
of 1.3--1.5 (M09), one can detect about $\sim$37--64 times
more foreground sources than the central bulge sources with the same flux limit.  
Combining these two factors, we should detect 0.013--0.022 foreground
sources for every central bulge source.
If we assume that about 10\% of the \nustar sources are AGN (see \S\ref{s:lnls}), 
63 out of the 70 \nustar sources found in the main GC region are near the GC. 
Then we expect to detect about 0.8--1.6 foreground sources among the
\nustar detections.
This is statistically consistent with no \nustar detection of foreground stars
in the survey.
Therefore, it is premature to conclude whether there is a population
difference of the \nustar selected X-ray sources between the GC and
other Galactic plane fields or alternatively whether there is a spectral
transition in between 10\sS{33} \lcgs at 8 kpc and 10\sS{32} \lcgs at
$\lesssim$ 2 kpc in the X-ray population. 

On the other hand,
M09, \citet{Hong09} and \citet{Heard13} suggest a possible enhancement
in the soft ($<$10 keV) X-ray source population relative to the stellar distribution near the
GC \footnote{In the case of \citet{Heard13}, the claim was
made for a lower luminosity limit ($\sim$10\sS{31} \lcgs) than this
survey or the other two studies ($\sim$10\sS{32} \lcgs).}.
In addition, they also show a possible spectral difference between the central
X-ray source population and the local field population, i.e.~the average photon
index ($\Gamma$\Ss{S}) of the bulge sources is closer to $\sim$1, which appears harder
than accreting foreground sources.  The \chandra results, however, are not
conclusive due to the degeneracy between the high absorption and
the intrinsic hardness of the X-ray spectrum in the \chandra energy band \citep[see
\S\ref{s:lnls} and ][]{Hailey16}.  The broadband coverage by \nustar 
can break this degeneracy and address this issue somewhat
conclusively, but it requires a large increase in the survey depth
or area. 
\citet{Fornasini16} have identified three foreground \nustar sources in
the Norma region, which is still statistically consistent with our results
due to low statistics.  The survey area can be effectively extended by
collecting serendipitous\footnote{For a fair comparison,
the target of each observation should not be counted unless the observation
is a part of a blind survey.} foreground
\nustar detections or measuring the lack thereof from other \nustar
observations of Galactic plane fields in the future.


\begin{figure} \begin{center}
\includegraphics*[width=0.35\textwidth,clip=true] {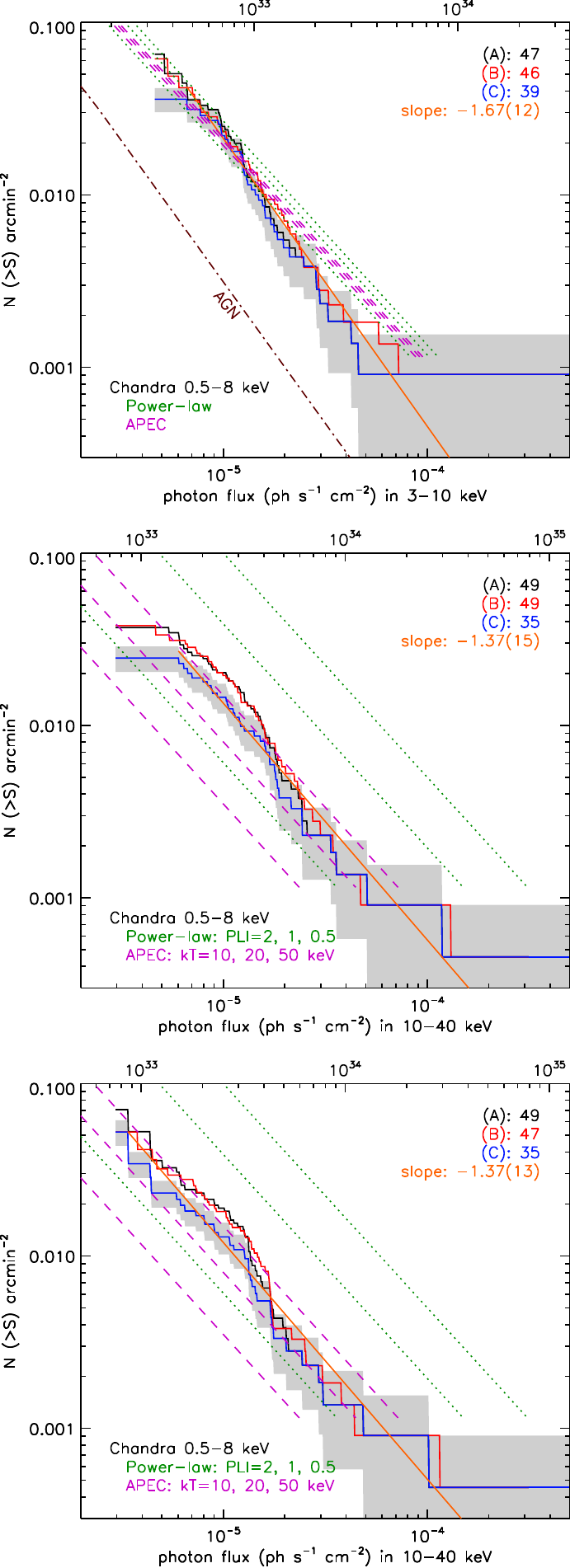}
\caption{Comparison of the \nustar 3--10 keV  (top) and
10--40 keV \lnls distributions (middle \& bottom) with the
\chandra 0.5--8 keV \lnls distribution. 
The top two panels use the model-based flux estimation, and the bottom panel use
the non-parametric flux estimation (\S\S\ref{s:flux} \& \ref{s:lnls}).
(A-black) The group-1 and 2 sources with the first
aperture sets (see \S\ref{s:aperture}), (B-red) with the second aperture
sets and (C-blue) the group-1 sources with the first aperture
sets.
The grey region shows the statistical errors for (C) and the
(yellow) solid line is the best fit for (C).  
The (green) dotted and (magenta) dashed lines show the \chandra 0.5--8 keV
\lnls distribution from M09: green dotted
lines are for power-law models with $\Gamma$ = 2, 1 and 0.5 (from bottom
to top) and magenta dashed lines are for APEC with $kT$ = 10, 20 and
50 keV (from bottom to top). The brown dashed-dot line in the top
panel is AGN \citep{Kim07}.
}
\label{f:lnls}
\end{center}
\end{figure}

\subsection{LogN-LogS Distributions} \label{s:lnls}

Fig.~\ref{f:lnls} shows the \lnls distributions of the \nustar sources in
the 3--10 (top) and 10--40 keV (middle and bottom) bands.  
The lower $x$-axis is the
observed photon flux ($F_X$) and the upper axis the observed X-ray luminosity ($L_X$) in the
same band at 8 kpc.
The conversion factor between the two $x$-axes is an average value of
the conversion factors for all the sources.  
Specifically, $F_X$~$\sim$~10\sS{-5} \pcgs corresponds to
$L_X$~$\sim$~\nep{7.1}{32} \lcgs in the 3--10 keV band and
$L_X$~$\sim$~\nep{2.5}{33} \lcgs in 10--40 keV.
The top panel also shows the AGN distribution using a photon index of 1.7
with \nH~=~1.2$\times$10\sS{23} cm\sS{-2} based on the
extragalactic
survey results by \citet{Kim07}. We expect about 10\% of the \nustar detections
to be AGN.

In order to estimate the systematic errors, we compare the distributions
prepared in three different methods: case (A) uses both group-1 and 2
sources with the photon fluxes based on the the first aperture sets,
case (B) is the same but with the second aperture sets, and case (C) is for only
the group-1 sources with the first aperture sets.  The largest difference
among the three cases in each flux bin is considered to be the systematic
errors of the \lnls distributions.  The statistical errors for case (C)
are shown in the grey shade in Fig.~\ref{f:lnls}.
A logarithmic linear fit to case (C) is shown by the yellow solid line,
which follows a relation of $N(>S) = N_0 (S/S_0)^{-\alpha}$. 
For comparison, we also overplot the the
\chandra\  0.5--8 keV \lnls distribution (M09) scaled to the
\nustar bands for six different spectral models: $\Gamma$ = 0.5,
1 and 2 for absorbed power-law models and $kT$ = 10, 20 and 50 keV for
single temperature APEC models. Note that M09 assumed $\Gamma$ = 0.5.  
We assume \nH = 6$\times$10\sS{22} cm\sS{-2} for all the models above.
Since all of the \nustar detections
have \chandra counterparts or at least candidates, both
\chandra and \nustar distributions should be consistent with each other.

The \nustar 3--10 keV \lnls distribution has a slope of $\alpha$ $\sim$
1.7 $\pm$ 0.1. It shows some deficit relative to the \chandra distribution
since the \nustar detections did miss some bright \chandra
sources likely due to on-going variability but they match within
1.5$\sigma$ of the statistical uncertainties.
It is clear that the soft energy band below 10 keV is insensitive to
the assumed spectral models in converting the \chandra distribution,
in part due to the similarity between the two bands in comparison given the absorption
towards the GC (i.e.~the fluxes below 2 keV do not contribute much).

In the 10--40 keV band, the luminosity distribution of the \nustar
sources shows a slope of 1.4 $\pm$ 0.1. 
Unlike the 3--10 keV band, in 10--40 keV, the assumption of the average
spectral shape in translating the 0.5--8 keV \chandra distribution
makes a significant difference.  For an absorbed power-law model,
the average photon index $\Gamma$ should be somewhere in between 1.5 and 2 or
for an absorbed single-temperature APEC model,
the average temperature $kT$ should be somewhere in between 20 and 50 keV in order
for the two distributions to match. This result is also consistent with
the photon index distribution in Fig.~\ref{f:dist}.

Since the flux of each source in the \lnls distribution (the middle
panel in Fig.~\ref{f:lnls}) is calculated for a power-law model with
the median energy-based photon index used in Fig.~\ref{f:dist}, one can argue that
the consistency in the overall photon index distribution between the two
figures may not be the result of two entirely independent analyses.
For a sanity check, we re-derive the \lnls distribution using
the flux values that are calculated
non-parametrically and model-independently (\S\ref{s:flux}),
as shown in the bottom panel in Fig.~\ref{f:lnls}. 
The non-parametric flux estimation results in a bit lower
flux values for the faint sources. This is in part because dividing
a relatively small number of X-ray counts from faint sources into
each small energy bin of 1 keV step can lead to some signal loss in
the non-parametric calculation.  On the other hand, the model-based
calculation tends to overestimate the flux of the faint sources depending on
how accurately the assumed model represents the true X-ray spectrum of
each source. Regardless of some differences near the faint end,
the \nustar 10--40 keV \lnls distributions of both model-based and
model-independent fluxes require the similar average spectral types for X-ray emission
of the \nustar sources in order to be consistent with
the \chandra distribution.

With the \chandra energy band alone, it is hard to constrain the X-ray
spectra of the GC X-ray sources, but the \nustar observations put a tighter
constraint on the plasma temperature for a thermal model or photon
index.\footnote{By ``photon indices", we mean an equivalent photon index for a simple power-law model. As
seen with the cut-off power-law model (the right panel of
Fig.~\ref{f:qd}), the photon indices can be easily skewed with additional
parameters when a more complex model is employed.  The soft band ($<$
10 keV) photon index ($\Gamma_s$) of $\sim$ 0.5 observed by \chandra can
be consistent with the broadband $\Gamma$ of 1.5 observed by \nustar if
there is a high energy cut-off at around 10 keV in the power-law spectrum.}

\subsection{Are MCVs Dominant in \nustar X-ray
Sources?} \label{s:mcv}

Of nine relatively bright \nustar sources in \S\ref{s:bright}, four are
NS X-ray binary systems, one is a run-away pulsar, another is suspected
to a BH or NS X-ray binary and the other three are suspected to be MCVs.
Of 15 \nustar sources with their \chandra spectra
model-fitted to search for the iron lines (\S\ref{s:fit}),
11 sources show a sign of the iron lines, seven of which have the broadband
photon index $\Gamma$~$<$~1.5, whereas two in the other
four sources without the iron lines have $\Gamma$~$<$~1.5.  
Among the combined 24 sources,
about 70\% of the relatively hard sources with $\Gamma$~$<$~1.5 show the 
iron lines, whereas only about 40\% of the relatively soft sources with
$\Gamma$~$>$~1.5 show the iron lines.  
The iron lines, combined with a hard
continuum ($\Gamma$~$\lesssim$~1.5 or $\Gamma$\Ss{S}~$\lesssim$~1 for
an absorbed power-law model), is a good indicator of a MCV rather than
a NS or BH X-ray binary.  The relatively high percentage of the iron
lines\footnote{A typical EW of the iron lines from MCVs
ranges from $\sim$150 to 300 eV \citep{Ezuka99}. On the other hands,
quiescent XBs ($\lesssim$ 10\sS{33} \lcgs) do not appear to exhibit detectable
Fe line emission ($\lesssim$ 50--120 eV) although their sample
size is small \citep[e.g.][]{Bradley07,Chakrabarty14,Rana16}.} 
among the sources with the hard continuum ($\Gamma$~$\lesssim$1.5)
indicates that the \nustar source population contain a large fraction of
MCVs, at least 40\% of the above 24 sources. Or if we consider all the sources
with the iron lines as MCVs, the fraction increases to $\sim$ 60\%. Then
roughly the other 40\% can be BH or NS X-ray binaries. 

On the other hand, \citet[][2012]{Degenaar10} detected 17 transients within the central
1.2 deg\sS{2} of the GC based on long term monitoring programs of the
GC region using \chandra and \swift.  These programs cover more or less
the complete sample of the BH or NS X-ray binaries with recurring bursts
on a time scale of less than a decade.  Among 14 in our survey field,
four were detected by \nustar
(Table~\ref{t:ap}), seven were unresolved in the Sgr A diffuse
complex, and the other three were undetected.
The relatively small number of transients observed in the GC region
indiciate that the fraction of BH or NS X-ray binaries is likely much
smaller than 40\%.

As seen in Fig.~\ref{f:dist}, the 3--40 keV luminosities at 8 kpc of the
\nustar sources are mostly in a range of 10\sS{33-34} \lcgs, where 
both quiescent NS or BH X-ray binaries and bright MCVs can be found.  The
broadband spectral properties of the \nustar sources show that both of
these types can contribute significantly to the \nustar source population.
We expect that the relative fraction of MCVs in the remaining
fainter \nustar sources can be much higher as their luminosity range falls
into a more typical luminosity range of IPs. Therefore, the overall fraction of 
MCVs in the \nustar sources is expected be greater than $\sim$ 60\%.

A key result of our survey is that the hard X-ray spectra of the
\nustar sources in the GC region are consistent with the apparent diffuse,
central hard X-ray emission (CHXE) found by
\citet{Perez15}. A leading scenarios is that the diffuse hard X-ray
emission is from 1,000 -- 10,000 unresolved IPs with high mass WDs,
which can produce high temperature plasma above 30 keV.  For a single
temperature model, such a high temperature translates to WD masses of
$\gtrsim$ 0.8\Ms, which is much higher than the average WD mass of
$\sim$ 0.6 \Ms in MCVs that are suspected to be responsible for the
Galactic Ridge X-ray emission (GRXE).  
For instance, a broadband (2--50 keV) analysis of the GRXE
from the \suzaku observations of
the Galactic Bulge within 1--3\Deg of the GC
also shows a lower plasma temperature (12--15 keV)
for the overall combined X-ray spectra \citep{Yuasa12}.
Bright MCVs found in the Norma region by
\citet{Fornasini16} also exhibit a noticeably lower plasma temperature for a
single temperature model ($<$ 20 keV in the Norma region vs.~$>$20 keV
in the GC region).  Note hard X-ray CVs selected by
\integral/IBIS in the field do show an average temperature
of $kT$~$\sim$22 keV \citep{Landi09}, but given the limited band width on
the soft X-ray side ($\gtrsim$ 15 keV), CVs detected by \integral/IBIS or
\swift/BAT likely have a selection bias toward high plasma temperature,
wheres the boardband coverage by \nustar is relatively free of such
a bias.

A possible scenario resulting in high mass ($>$ 0.8 \Ms) WDs in the
CVs near the GC is that the GC region harbors a
large number of $>$ 4 \Ms B-stars, 
compared to the field, given the WD initial-final mass relation according to
\citet{Andrews15}.
\citet{Hailey16} argue that the excess B-star population
needed to explain high WD masses is within the large uncertainty
of the expected population in the GC region.
On the other hand, the average mass of the WD in the
non-magnetic CVs or isolated magnetic WDs are about 0.8\Ms
\citep{Wijnen15,Ferrario15}.
In addition, there is no clear evidence for high mass progenitors
for the WDs in CVs \citep{Zorotovic11}.  Since the highly magnetized,
isolated WDs are considered to be products of binary evolution, perhaps
the binary evolution may be responsible for high mass WDs in the CVs.
Then the relatively low WD mass from the X-ray observations of the
field is more unusual than the projected high mass of the WD based on
the X-ray spectral analysis of the sources in the GC region.

The similarity in the broadband X-ray spectra of the CHXE
and the \nustar sources in this survey reinforces the
scenario that (1) the X-ray population in the GC region is predominantly
MCVs but also with a significant fraction of NS and BH X-ray binaries,
and (2) the GC region also harbors an increasingly higher fraction of MCVs
with high WD masses that produce harder X-rays than those in other
regions in the plane.

\subsection{MSPs or Young Pulsars in \nustar X-ray Sources?} \label{s:msp}

Another interesting proposal for the CHXE by
\citet{Perez15} is that it can be the result of the unresolved non-thermal
emission from a large population of millisecond pulsars (MSPs).
To explain the total observed luminosity of \nep{2}{34} \lcgs with
rotationally powered systems, about 4000 MSPs would be needed with an
average non-thermal X-ray ($L$\Ss{n,X}) luminosity of \nep{5}{30}
\lcgs under the assumption of $L$\Ss{n,X}$\sim$10\sS{-4}$\dot{E}$
according to \citet{Takata12}, where $\dot{E}$ is the spin-down power.

The recent \fermi observations of excess gamma-ray emission
in the inner galaxy \citep{Goodenough09, Hooper11} triggered a series
of debates regarding its origin: e.g.~dark matter annihilation
\citep[e.g.][]{Hooper11b} or a collection of unresolved MSPs
\citep[e.g.][]{Abazajian12} or young pulsars \citep{OLeary15}. 
\citet{Lee15} presented evidence of unresolved gamma-ray point sources
in the \fermi observations of the inner galaxy by demonstrating that a
simple pure Poisson distribution is inadequate to explain the observed
distribution of the excess gamma-ray photons within the central few
degrees. If gamma-ray point sources are required to explain the excess,
MSPs and young pulsars become the leading candidates given their dominance
in the Galactic \fermi source population.

According to \citet{Hooper13}, the number of the MSPs required to explain
the gamma-ray excess exceed by a factor of 10 what is projected from
the observed field population.  \citet{Cholis14} also argue against
MSPs as the source of the gamma-ray excess, based on the paucity of
the resolved sources within the central 10\Deg of the
GC.  Contrarily \citet{Brandt15} proposed the Galactic Bulge as a
giant collection of disrupted globular clusters, which can naturally
lead to an enhancement of MSPs and subsequently explain the excess
gamma-ray emission.  In fact, \citet{Hooper13} pointed out that the
inner tens of parsecs from the GC could have high MSP population as
massive globular clusters, and thus the model by \citet{Brandt15}
effectively extends the region with a high population of MSPs to a few
kpc scale ($\sim$10\Deg).  On the other hand, \citet{OLeary15} argue
that the excess gamma-ray emission can be explained with a reasonable
number of young pulsars, given their relatively bright gamma-ray
emission \citep{Abdo13}. 
However, it appears difficult to explain the
drastic difference in scale and morphology between the CHXE
(asymmetric, parsec scale) and
the excess gamma-ray emission (symmetric, kpc scale) with a single
type of the source population, either MSPs or young pulsars.  

In the case of the \nustar X-ray sources in our survey,
the 10--40 keV X-ray luminosities for the majority are $\sim$10\sS{33-34}
\lcgs. It is not straightforward to estimate the gamma-ray luminosity
($L_\gamma$) of rotationally powered pulsars from its X-ray luminosity
($L_X$), but if we assume that the hard X-ray emission from the \nustar
X-ray sources is a non-thermal component of rotationally powered pulsars,
we can roughly estimate $L_\gamma$ through
the spin-down power \citep{Marelli11}, with the relation
$L$\Ss{n,X}$\sim$10\sS{-4}$\dot{E}$.  Then, the expected $L_\gamma$
for many of the \nustar sources exceeds 10\sS{35-37}~\lcgs, which
corresponds to 10\sS{-11}--10\sS{-9} \fcgs at 8 kpc.

According to the \fermi pulsar catalog \citep{Abdo13}, the brightest MSPs
have $L_\gamma$ $\sim$ 10\sS{34} \lcgs, and the \nustar sources
are much too bright to be the typical MSPs found by \fermi.  Young pulsars are
brighter than MSPs, but the expected $L_\gamma$ of the \nustar sources
are still near or above the brightest young pulsars observed by \fermi. 
Since the 50\% completeness limit of individual source detection at the GC 
is about \nep{4}{-11} \fcgs according to \citet{OLeary15}.  the expected
$L_\gamma$ of the brightest \nustar sources is large enough for \fermi to
resolve individually. Although the \nustar sources in this survey would
be likely spatially confused for \fermi, similar hard
X-ray sources in the vicinity of the region, if they are rotationally powered 
pulsars, could have been
resolved by \fermi as suggested in \citet{Cholis14}.
In addition, at least some young pulsars are expected to be associated
with PWNe with observable soft ($<$ 10 keV) X-ray filaments, depending on their ages
\citep[e.g.][suggested young pulsars created in the last 300
kyr based on 34 X-ray filaments]{Muno08}, but the majority of the \nustar sources do not
show any association with the soft ($<$10 keV) X-ray filaments.
Therefore, at least the bright \nustar X-ray sources in our catalog are
believed not to be typical rotationally powered pulsars
unless the \nustar X-ray sources are unusual pulsars with
much higher-than-usual X-ray luminosities for the given spin-down power.

\section{Summary and Future Work}

\begin{compactenum}
\item  We have discovered 70 hard X-ray sources in the 0.6 deg\sS{2}
region around the GC and 7 in the Sgr~B2 cloud field. Of the 77 sources,
66 sources show significant X-ray emission in hard ($>$ 10 keV) X-ray bands.
\item  The broadband (3--40 keV) energy quantiles of the \nustar sources show that for a
power-law model the majority of the sources have photon indices of $\Gamma$ = 1--2 and
about 20\% with $\Gamma$ $<$ 1.
\item  The 3--10 keV \lnls distribution of the \nustar sources is in
a good agreement with the 0.5--8 keV \chandra distribution of the GC region. 
\item  The \nustar 10--40 keV and \chandra 0.5--8 keV \lnls distributions
match if  the average photon index ($\Gamma$) of the \nustar sources is in between 1.5 and 2
for a power-law model or the plasma temperature lies between 20 and 50 keV for
a single temperature APEC model. 
\item  For an absorbed power-law model, the average soft ($<$ 10 keV)
band photon index ($\Gamma$\Ss{S}) of
the \chandra sources in the GC region was estimated to be 0.5--1
\citep[M09;][]{Hong09}, which is smaller than the broadband (3--40
keV) photon
index ($\Gamma$) measured here for the \nustar sources.
The limited \chandra energy band is responsible
for the discrepancy, but 
if the X-ray spectra of
the GC region X-ray sources have an exponential cut-off at
$\sim$ 10 keV, the apparent photon indices of the \chandra and \nustar
spectra match.
\item The spectral analysis of the relatively bright 24 sources suggests that 
MCVs comprise $>$ 40--60\% of the total, and NS or BH
X-ray binaries can make up the rest.
The fraction of MCVs among the fainter sources is likely higher ($>$60\%).
\item The \nustar sources in the GC region exhibit
higher plasma temperatures than the hard X-ray sources in the field
including the Norma region by \citet{Fornasini16}.  If MCVs 
comprise a large majority of the \nustar sources in the GC region,
the observed plasma temperature
range translates to a WD mass of $\gtrsim$ 0.8 \Ms, which is higher than
the field average of $\sim$ 0.5 \Ms \citep{Hailey16}.
\item  A large population of IPs with higher mass WDs can
explain the average X-ray spectrum of the \nustar sources in 
this survey as well as the CHXE discovered by \citet{Perez15}.   
\item If the \nustar X-ray
sources in the GC region are rotationally
powered pulsars, their expected $L_\gamma$ exceeds typical values of
both MSPs and young pulsars observed by \fermi. Therefore, the \nustar
X-ray sources in this survey do not likely contain many rotationally
powered pulsars.
\item  The \nustar detections lack foreground sources, which is
significantly different from the \chandra source population but
it is still consistent with the stellar population, given the
sensitivity and coverage limitations of the present survey.
\end{compactenum}

It is essential to continue monitoring of the GC region for understanding
the nature of the hard X-ray sources and the GC region as a whole.
To acquire broadband X-ray spectra of several \nustar sources with high
photon statistics, deep exposures of selected regions in block A
is planned under the \nustar legacy program.

\section{Acknowledgment} 

This work was supported under NASA Contract No.  NNG08FD60C, and made
use of data from the \nustar mission, a project led by the California
Institute of Technology, managed by the Jet Propulsion Laboratory, and
funded by the National Aeronautics and Space Administration.  We thank
the \nustar Operations, Software and Calibration teams for support with
the execution and analysis of these observations.  We thank G.~Ponti for
careful reading and suggestions of the manuscript.  This research has made
use of the \nustar Data Analysis Software (NuSTARDAS) jointly developed by
the ASI Science Data Center (ASDC, Italy) and the California Institute
of Technology (USA).  J.~Hong acknowledges support from NASA/APRA
grant NNX14AD59G.  R. Krivonos acknowledges support from Russian
Science Foundation through grant 14-22-00271.  F.E. Bauer acknowledges
support from CONICYT-Chile (Basal-CATA PFB- 06/2007, FONDECYT 1141218,
"EMBIGGEN" Anillo ACT1101), and the Ministry of Economy, Development,
and Tourism's Millennium Science Initiative through grant IC120009,
awarded to The Millennium Institute of Astrophysics, MAS. S. Zhang is
supported by NASA Headquarters under the NASA Earth and Space Science
Fellowship Program - Grant "NNX13AM31". D. Barret acknowledges support
from the French Space Agency (CNES).

\section{Appendix} 
\label{s:tm}

\subsection{Resolving Power of the \nustar Optics} \label{s:resolve}


\begin{figure} \begin{center}
\includegraphics*[width=0.48\textwidth,clip=true] {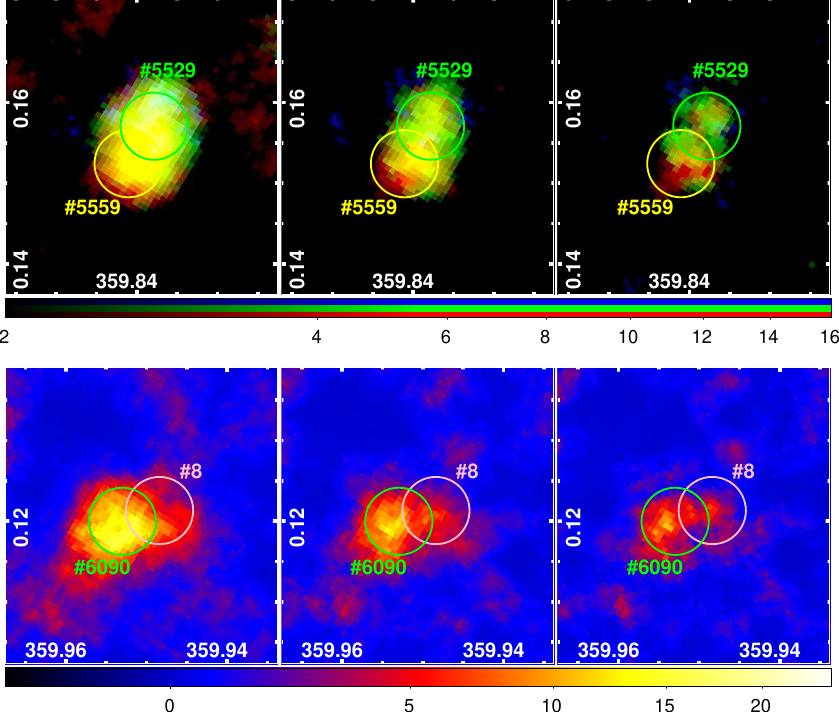}
\caption{Resolving CXOUGC J174437.1--285934 (\chandra ID \#5529 or NGP 32) and
CXOUGC J174438.7--285933 (\#5559 or NGP 54)
(top) and GRS 1741.9--2853 (\#6090 or NGP 2) vs.~CXOUGC J174501.3--285501 (\#8) (bottom).
The images are the trial maps of
30\%, 20\% and 15\% PSF enclosures from left to right.
The top panels are the three color trial maps (red: 3--10 keV, green:
10--20 keV, blue: 20--40 keV) and 
the bottom panels are in 20--40 keV.
The circles show 15\arcsec\ radii around the \chandra source positions.
}
\label{f:5529}
\end{center}
\end{figure}

Two \chandra sources, CXOUGC J174437.1--285934 and CXOUGC J174438.7--285933, 
are located about 20\arcsec\ apart from each other and show
similar photon fluxes in the 2--8 keV band (M09).  In \nustar,
the three-color trial maps in the same region show two bright spots
(NGPs 32 and 54) with distinct X-ray colors,
which are also separated by about 20\arcsec\ (top panel in Fig.~\ref{f:5529}).
The separations between these two \nustar spots and 
the \chandra sources are about 5--6\arcsec. 
Based on the relative proximity to the \chandra sources,
we associate NGP 32 to CXOUGC J174437.1--285934 and NGP 54 to CXOUGC J
174438.7--285933.

Five observations covered the region: two observations do not have
bright sources to define a clear astrometric correction, and the other
three require $<$5\arcsec\ boresight shifts. Since the boresight
shifts are mostly less than 10\arcsec\ (Fig.~\ref{f:bs}), we believe
that the 20\arcsec\ separation between the two spots in the trial map
is too large for an astrometric error.  In addition, trial maps made of
the three observations with astrometric corrections show similar results
(not shown).  On the other hand, the largest offset
between the \nustar and \chandra sources in Table~\ref{t:src} is about
13.5\arcsec, and thus it is very unlikely but possible that the
combination of a large astrometric error and a large statistics-driven
positional uncertainty may produce an artificial 20\arcsec\ separation.  
If so, then the source must have been experiencing a remarkable spectral
variation (see below).  Considering these factors, we assign the
brighter of the two in group 1 and the other in group 2.

The \nustar quantile diagram shows that the X-ray spectrum of NGP 32 is
harder and more consistent with a power-law model while NGP 54 is
softer and more consistent with a thermal model (Fig.~\ref{f:qd}). 
The \chandra quantile diagram (Fig.~\ref{f:qd_chandra} in
\S\ref{s:missing}) also shows that  CXOUGC J174437.1--285934  has a harder
X-ray spectrum ($\Gamma$\Ss{S} $\sim$ 1) than CXOUGC J174438.7--285933
($\Gamma$\Ss{S} $\sim$ 3). 

The 20--40 keV trial map shows another similar case of
two nearby sources (GRS 1741.9--2853 vs.~CXOUGC J174501.3--285501, 17\arcsec\  apart). Since 
GRS 1741.9--2853 is predominantly brighter at low energies below 20 keV,
an additional peak is only resolved in
the 20--40 keV trial map (the bottom panel in Fig.~\ref{f:5529}, too
faint above 40 keV).  The significance of the new spot in the trial map
appears to be roughly consistent with the 2--10 keV flux differences
between the two \chandra sources (a factor of $\sim$ 20 according to
M09). The burst-only data of GRS 1741.9--2853 do not show any peak with
a similar significance  near CXOUGC J174501.3--285501, implying that
the peak in the survey trial maps is not likely an artifact of the
large PSF wings of GRS~1741.9--2853. On the other hand, there is no clear sign
of CXOUGC J174501.3--285501 in the \nustar data during the quiescent
period of GRS~1741.9--2853.  Our re-analysis of the \chandra archival
data does not show any significant sign of X-ray signals at the location
of CXOUGC J174501.3--285501, which questions the validity of the \chandra
detection.  Without the \chandra counterpart, a marginal detection near
the threshold in one energy band technically does not meet our source
selection criteria.  Therefore, we excluded   CXOUGC J174501.3--285501 as
part of the \nustar detections.  Confirmation of CXOUGC J174501.3--285501
as a real detection will require additional \chandra observations with GRS
1741.9-2853 near the aimpoint when GRS~1741.9--2853 is relatively faint.

\subsection{Diffuse Emission}\label{s:diffuse}

The trial maps generated with fixed-size detection cells
retain diffuse emission structures in convolution with the PSF
at the scales of the cell sizes.
The prominent diffuse structures seen in the trial maps include the
Sgr~A complex,
a few molecular clouds \citep{Mori15} and X-ray filaments 
including G359.89--0.08 \citep[Sgr~A-E;][]{Zhang14}, the Arches
Cluster \citep{Krivonos14}, G359.97--0.038 \citep{Nynka15}, and the Sgr~B2
cloud \citep{Zhang15}.
These regions are excluded in our point source analysis.
Since the trial maps show the emission significance in general,
it would require iterative forward modeling with a proper emission morphology to
extract the flux or intensity of the emission structure from a trial
map, which is beyond the scope of this paper.  On the other hand, 
the trial maps reveal a general trend of emission morphology more
clearly than raw images.

The three color trial map of the Arches cluster (the middle panel in
Fig.~\ref{f:sgra}) shows an elongated bright hard X-ray streak embedded in
a soft circular diffuse structure \citep[see a wavelet analysis
in][]{Krivonos14}. In the MC1 region, the trial map 
(the bottom panel in Fig.~\ref{f:sgra}) shows a possible
spatial separation between the soft and hard X-ray emission. The
soft X-ray emission is more extended along Galactic latitude and
closer to Sgr~A*, whereas the hard X-ray emission is more central with
respect to the Galactic plane and further away from Sgr~A*.  The separation
is too small (about 7\arcsec) to rule out systematic artifacts, but it
does support the idea of a morphological spectral variation
in the region. For instance, \citet{Clavel13} claimed a detection of spatial 
variation in the evolution of the Fe K\Ss{\alpha} line and
argued for an X-ray reflection nebula model \citep[see also][]{Ponti13} 
where the X-ray emission from the cloud was
triggered by bright flares of Sgr~A* in the past. See \citet{Mori15}
for an in-depth analysis of the \nustar observations of MC1.


\begin{table*}
\scriptsize
\begin{minipage}{0.99\textwidth}
\caption{\nustar Observations of the Galactic Center Region}
\begin{tabular*}{\textwidth}{r ccc crrc @{} R rrl}
\hline\hline
                          & \multicolumn{2}{c}{                } & (1)    &                &       & (2)   & (3)     & (4)             &             &           & (5)                    \\
Observation               & \multicolumn{3}{c}{Pointing (J2000)} & Start Time     & Exp.  & GTI   & SL      & \multicolumn{4}{c}{Boresight Shift} \\
\cline{2-4}\cline{9-12}ID & R.A.             & Dec      & PA     &                &       &       & Removal & Total           & \ \ in R.A. & in Dec    & Reference Source       \\
                          & (\Deg)           & (\Deg)   & (\Deg) & (UT)           & (ks)  & (ks)  & (FPM)   & (\arcsec)       & (\arcsec)   & (\arcsec) & (IDs)                  \\

\hline \multicolumn{11}{l}{\it (Sgr A*)} \\ 
30001002001               & 266.4168         & -29.0078 & 333    & 12-07-20 02:11 & 167.3 & 122.1 & B       & 5.89            & 6.12        & -2.45     & 2743, 1568, 2331       \\
02003                     & 266.4168         & -29.0078 & 336    & 12-08-04 07:56 & 83.5  & 80.4  & B       & 5.15            & 5.15        & -2.51     & 2743, 6090, 1568, 2331 \\
02004                     & 266.4168         & -29.0078 & 343    & 12-10-16 18:31 & 53.9  & 52.4  & B       & 5.89            & 6.12        & -2.45     & 2743, 1568, 2331       \\
\hline
\end{tabular*}

\label{t:obs}
Notes. (1) Pointing angle. 
(2) Sum of good time intervals. The data of
Sgr~A* flares were excluded. (3) Focal plane module (FPM) where stray
light background photons from nearby bright sources were removed.
(4) Applied boresight shift. 
(5) Reference \chandra sources used for boresight shift: IDs are from M09.
\end{minipage}
\end{table*}

\begin{sidewaystable}
\scriptsize
\vspace{10cm}
\caption{\nustar Galactic Point (NGP) Source List
}
\begin{tabular*}{\textwidth}{c cc ccr rr c rr rr l}
\hline\hline
   (1)                               & \multicolumn{2}{c}{(2)             } & (3)                  & (4)             & (5)    & (6)       & (7)  & (8) & (9)        & (10)   & (11) & (12) & (13)                                \\
No.                                  & \multicolumn{2}{c}{\nustar Position} & \multicolumn{3}{c}{\chandra Counterpart} & Offset    & Exp. & Det & No.        & Energy & Enc. & No.  & Comments                            \\
\cmidrule(r){2-3}\cmidrule(l){4-6}ID & R.A.             & Dec       & ID                   & Name            & Flux   &           &      &     & Trials     & Band   & Size & Det. &                                     \\
(NGP)                                & \multicolumn{2}{c}{(\Deg, J2000)   } &                      & (CXOUGC J)      & (cgs)  & (\arcsec) & (ks) &     & (10\sS{X}) & (keV)  & (\%) &      &                                     \\

\hline \multicolumn{14}{l}{\it Group 1} \\ 
1                                    & 266.58731        & -28.72770 & 7722                 & 174621.0-284342 & 7330.0 & 3.7       & 26   & SH  & 14186.0    & 3-79   & 30   & 15   & 1E 1743.1-2843                      \\
2                                    & 266.26047        & -28.91413 & 6090                 & 174502.3-285449 & 3330.0 & 2.1       & 301  & SH  & 6881.1     & 3-10   & 30   & 15   & GRS 1741.9-2853                     \\
3                                    & 266.05738        & -28.90679 & 4942                 & 174413.7-285423 & 29.5   & 0.4       & 208  & SH  & 89.4       & 3-79   & 30   & 15   &                                     \\
\hline
\end{tabular*}

\label{t:src}
\end{sidewaystable}

\addtocounter{table}{-1}
\begin{sidewaystable*}
\scriptsize
\vspace{6cm}
\caption{\nustar Galactic Point (NGP) Source List: continued from the previous page}
\begin{tabular*}{\textwidth}{c cc ccr rr c rr rr l}
\hline\hline
   (1)                               & \multicolumn{2}{c}{(2)             } & (3)                  & (4)             & (5)   & (6)       & (7)  & (8) & (9)        & (10)   & (11) & (12) & (13)                                      \\
No.                                  & \multicolumn{2}{c}{\nustar Position} & \multicolumn{3}{c}{\chandra Counterpart} & Offset    & Exp. & Det & No.        & Energy & Enc. & No.  & Comments                                  \\
\cmidrule(r){2-3}\cmidrule(l){4-6}ID & R.A.             & Dec       & ID                   & Name            & Flux  &           &      &     & Trials     & Band   & Size & Det. &                                           \\
(NGP)                                & \multicolumn{2}{c}{(\Deg, J2000)   } &                      & (CXOUGC J)      & (cgs) & (\arcsec) & (ks) &     & (10\sS{X}) & (keV)  & (\%) &      &                                           \\

\hline \multicolumn{14}{l}{\it Group 2 (unclear association)} \\ 
53                                   & 266.59263        & -28.88025 & 7732                 & 174621.5-285256 & 6.9   & 11.6      & 121  & SH  & 37.0       & 3-79   & 30   & 12   & \#7757 (41\arcsec, 5.1x)                  \\
54                                   & 266.16050        & -28.99325 & 5559                 & 174438.7-285933 & 6.4   & 4.1       & 356  & SH  & 25.6       & 3-79   & 20   & 11   & \#5529 (20\arcsec, 1.0x)                  \\
55                                   & 266.35586        & -29.00918 & 706                  & 174525.5-290028 & 5.6   & 4.9       & 624  & SH  & 14.4       & 3-79   & 15   & 6    & \#676 (8.4\arcsec, 0.5x)                  \\
\hline
\end{tabular*}

\label{t:src}
See \S\ref{s:catalog} for the column definitions.

\caption{\nustar Point Sources in the Sgr~B2 Field
}
\begin{tabular*}{\textwidth}{c cc ccr rr c rr rr l}
\hline\hline
   (1)                               & \multicolumn{2}{c}{(2)             } & (3)                  & (4)             & (5)   & (6)       & (7)  & (8) & (9)        & (10)   & (11) & (12) & (13)                         \\
No.                                  & \multicolumn{2}{c}{\nustar Position} & \multicolumn{3}{c}{\chandra Counterpart} & Offset    & Exp. & Det & No.        & Energy & Enc. & No.  & Comments                     \\
\cmidrule(r){2-3}\cmidrule(l){4-6}ID & R.A.             & Dec       & ID                   & Name            & Flux  &           &      &     & Trials     & Band   & Size & Det. &                              \\
(NGP)                                & \multicolumn{2}{c}{(\Deg, J2000)   } &                      & (CXOUGC J)      & (cgs) & (\arcsec) & (ks) &     & (10\sS{X}) & (keV)  & (\%) &      &                              \\

\hline \multicolumn{14}{l}{\it Group 1} \\ 
71                                   & 266.68683        & -28.26511 & 8095                 & 174645.2-281547 & 68.4  & 8.6       & 85   & SH  & 73.4       & 3-10   & 30   & 7    &                              \\
72                                   & 266.92700        & -28.37488 & 8943                 & 174742.4-282227 & 13.6  & 3.3       & 134  & SH  & 31.7       & 3-79   & 30   & 12   &                              \\
73                                   & 266.76484        & -28.35974 & 8392                 & 174703.5-282136 & 3.8   & 2.2       & 141  & SH  & 24.1       & 3-79   & 30   & 12   &                              \\
\hline
\end{tabular*}

\label{t:src:SgrB2}
See \S\ref{s:catalog} for the column definitions.
\end{sidewaystable*}

\clearpage

\begin{sidewaystable}
\scriptsize
\centering
\vspace{10cm}
\caption{\nustar Aperture Photometry Results of \nustar Sources}
\begin{tabular*}{\textwidth}{cc r@{}l r@{}l r@{}l r@{}l D{.}{.}{2.1} r@{}l D{.}{.}{4.1} r@{}l r@{}l @{\ } r@{}l r@{}l cc}
\hline\hline
(1)                                    & (2)  & \multicolumn{2}{c}{(3)      } & \multicolumn{2}{c}{(4)     } & \multicolumn{2}{c}{(5)       } & \multicolumn{2}{c}{(6)     } & \multicolumn{1}{c}{(7)}                   & \multicolumn{2}{c}{(8)   } & \multicolumn{1}{c}{(9)}                 & \multicolumn{2}{c}{(10)     } & \multicolumn{2}{c}{(11)      } & \multicolumn{2}{c}{(12)                   } & \multicolumn{2}{c}{(13)      } & (14)                       & (15)    \\
No.                                    & ID   & \multicolumn{2}{c}{Net      } & \multicolumn{2}{c}{Hardness} & \multicolumn{2}{c}{Median    } & \multicolumn{2}{c}{Quartile} & \multicolumn{1}{c}{\nH}                   & \multicolumn{2}{c}{Photon} & \multicolumn{5}{c}{$F_X$ (10\sS{-6} ph s\sS{-1} cm\sS{-2})} & \multicolumn{4}{c}{$L_X$ (10\sS{32} \lcgs)} & Ap.                        & Flag    \\
\cmidrule(r){14-18}\cmidrule(l){19-22} &      & \multicolumn{2}{c}{Counts   } & \multicolumn{2}{c}{Ratio   } & \multicolumn{2}{c}{$E$\Ss{50}} & \multicolumn{2}{c}{Ratio   } & \multicolumn{1}{c}{}                      & \multicolumn{2}{c}{Index } & \multicolumn{1}{c}{\chandra}            & \multicolumn{2}{c}{\nustar  } & \multicolumn{2}{c}{\nustar   } & \multicolumn{2}{c}{\nustar                } & \multicolumn{2}{c}{\nustar   } & Size                       &         \\
                                       &      & \multicolumn{2}{c}{3--40 keV} & \multicolumn{2}{c}{        } & \multicolumn{2}{c}{(keV)     } & \multicolumn{2}{c}{QR      } & \multicolumn{1}{c}{(10\sS{22} cm\sS{-2})} & \multicolumn{2}{c}{      } & \multicolumn{1}{c}{2--8 keV}            & \multicolumn{2}{c}{3--10 keV} & \multicolumn{2}{c}{10--40 keV} & \multicolumn{2}{c}{3--10 keV              } & \multicolumn{2}{c}{10--40 keV} & IDs                        &         \\

\hline \multicolumn{24}{l}{\it Group 1} \\
\hline
\end{tabular*}

\label{t:ap}
\end{sidewaystable}

\addtocounter{table}{-1}
\begin{sidewaystable*}
\scriptsize
\vspace{6cm}
\caption{\nustar Aperture Photometry Results of \nustar Sources: continued from the previous page}
\begin{tabular*}{\textwidth}{cc r@{}l r@{}l r@{}l r@{}l D{.}{.}{2.1} r@{}l D{.}{.}{4.1} r@{}l r@{}l @{\ } r@{}l r@{}l cc}
\hline\hline
(1)                                    & (2)  & \multicolumn{2}{c}{(3)      } & \multicolumn{2}{c}{(4)     } & \multicolumn{2}{c}{(5)       } & \multicolumn{2}{c}{(6)     } & \multicolumn{1}{c}{(7)}                   & \multicolumn{2}{c}{(8)   } & \multicolumn{1}{c}{(9)}                 & \multicolumn{2}{c}{(10)     } & \multicolumn{2}{c}{(11)      } & \multicolumn{2}{c}{(12)                   } & \multicolumn{2}{c}{(13)      } & (14)                       & (15) \\
No.                                    & ID   & \multicolumn{2}{c}{Net      } & \multicolumn{2}{c}{Hardness} & \multicolumn{2}{c}{Median    } & \multicolumn{2}{c}{Quartile} & \multicolumn{1}{c}{\nH}                   & \multicolumn{2}{c}{Photon} & \multicolumn{5}{c}{$F_X$ (10\sS{-6} ph s\sS{-1} cm\sS{-2})} & \multicolumn{4}{c}{$L_X$ (10\sS{32} \lcgs)} & Ap.                        & Flag \\
\cmidrule(r){14-18}\cmidrule(l){19-22} &      & \multicolumn{2}{c}{Counts   } & \multicolumn{2}{c}{Ratio   } & \multicolumn{2}{c}{$E$\Ss{50}} & \multicolumn{2}{c}{Ratio   } & \multicolumn{1}{c}{}                      & \multicolumn{2}{c}{Index } & \multicolumn{1}{c}{\chandra}            & \multicolumn{2}{c}{\nustar  } & \multicolumn{2}{c}{\nustar   } & \multicolumn{2}{c}{\nustar                } & \multicolumn{2}{c}{\nustar   } & Size                       &      \\
                                       &      & \multicolumn{2}{c}{3--40 keV} & \multicolumn{2}{c}{        } & \multicolumn{2}{c}{(keV)     } & \multicolumn{2}{c}{QR      } & \multicolumn{1}{c}{(10\sS{22} cm\sS{-2})} & \multicolumn{2}{c}{      } & \multicolumn{1}{c}{2--8 keV}            & \multicolumn{2}{c}{3--10 keV} & \multicolumn{2}{c}{10--40 keV} & \multicolumn{2}{c}{3--10 keV              } & \multicolumn{2}{c}{10--40 keV} & IDs                        &      \\

\hline \multicolumn{24}{l}{\it Group 2 (unclear association)} \\
\hline
\end{tabular*}

\label{t:ap}
See \S\ref{s:photometry} for the column definitions.

\caption{\nustar Aperture Photometry Results of Sources in the Sgr~B2 field}
\begin{tabular*}{\textwidth}{cc r@{}l r@{}l r@{}l r@{}l D{.}{.}{2.1} r@{}l D{.}{.}{4.1} r@{}l r@{}l @{\ } r@{}l r@{}l cc}
\hline\hline
(1)                                    & (2)  & \multicolumn{2}{c}{(3)      } & \multicolumn{2}{c}{(4)     } & \multicolumn{2}{c}{(5)       } & \multicolumn{2}{c}{(6)     } & \multicolumn{1}{c}{(7)}                   & \multicolumn{2}{c}{(8)   } & \multicolumn{1}{c}{(9)}                 & \multicolumn{2}{c}{(10)     } & \multicolumn{2}{c}{(11)      } & \multicolumn{2}{c}{(12)                   } & \multicolumn{2}{c}{(13)      } & (14)                       & (15) \\
No.                                    & ID   & \multicolumn{2}{c}{Net      } & \multicolumn{2}{c}{Hardness} & \multicolumn{2}{c}{Median    } & \multicolumn{2}{c}{Quartile} & \multicolumn{1}{c}{\nH}                   & \multicolumn{2}{c}{Photon} & \multicolumn{5}{c}{$F_X$ (10\sS{-6} ph s\sS{-1} cm\sS{-2})} & \multicolumn{4}{c}{$L_X$ (10\sS{32} \lcgs)} & Ap.                        & Flag \\
\cmidrule(r){14-18}\cmidrule(l){19-22} &      & \multicolumn{2}{c}{Counts   } & \multicolumn{2}{c}{Ratio   } & \multicolumn{2}{c}{$E$\Ss{50}} & \multicolumn{2}{c}{Ratio   } & \multicolumn{1}{c}{}                      & \multicolumn{2}{c}{Index } & \multicolumn{1}{c}{\chandra}            & \multicolumn{2}{c}{\nustar  } & \multicolumn{2}{c}{\nustar   } & \multicolumn{2}{c}{\nustar                } & \multicolumn{2}{c}{\nustar   } & Size                       &      \\
                                       &      & \multicolumn{2}{c}{3--40 keV} & \multicolumn{2}{c}{        } & \multicolumn{2}{c}{(keV)     } & \multicolumn{2}{c}{QR      } & \multicolumn{1}{c}{(10\sS{22} cm\sS{-2})} & \multicolumn{2}{c}{      } & \multicolumn{1}{c}{2--8 keV}            & \multicolumn{2}{c}{3--10 keV} & \multicolumn{2}{c}{10--40 keV} & \multicolumn{2}{c}{3--10 keV              } & \multicolumn{2}{c}{10--40 keV} & IDs                        &      \\

\hline \multicolumn{24}{l}{\it Group 1} \\
\hline
\end{tabular*}

\label{t:ap_sgrb2}
See \S\ref{s:photometry} for the column definitions.
\end{sidewaystable*}

\clearpage

\begin{table*}
\scriptsize
\begin{minipage}{0.99\textwidth}
\caption{Spectral Model Fit Results of Four Bright \nustar Sources (see \S\ref{s:fit} and Fig.~\ref{f:spec})}
\begin{tabular*}{\textwidth}{r l l l l l l l l}
\hline\hline
Source ID & \multicolumn{2}{c}{NGP 3} & \multicolumn{2}{c}{NGP 4} & \multicolumn{2}{c}{NGP 7} & \multicolumn{2}{c}{NGP 34} \\
chandra ID & \multicolumn{2}{c}{\#4942} & \multicolumn{2}{c}{\#6369} & \multicolumn{2}{c}{\#5908} & \multicolumn{2}{c}{\#1568} \\
\hline\multicolumn{1}{l}{\it Energy Band (keV)} \\
\nustar & \multicolumn{2}{c}{5--26} & \multicolumn{2}{c}{3--23} & \multicolumn{2}{c}{3--24} & \multicolumn{2}{c}{5--26} \\
\xmm & \multicolumn{2}{c}{2.5--10} & \multicolumn{2}{c}{2-10} & \multicolumn{2}{c}{2--10} & \multicolumn{2}{c}{4--10} \\
\chandra & \multicolumn{2}{c}{2--10} & \multicolumn{2}{c}{2-10} & \multicolumn{2}{c}{2--10} & \multicolumn{2}{c}{2--10} \\
\cmidrule(r){2-3}\cmidrule(r){4-5}\cmidrule(r){6-7}\cmidrule(r){8-9}Models & \multicolumn{1}{c}{Power-law} & \multicolumn{1}{c}{APEC} & \multicolumn{1}{c}{Power-law} & \multicolumn{1}{c}{APEC} & \multicolumn{1}{c}{Power-law} & \multicolumn{1}{c}{APEC} & \multicolumn{1}{c}{Power-law} & \multicolumn{1}{c}{APEC} \\
\hline\multicolumn{1}{l}{\it Relative Normalization} \\
\hline
\end{tabular*}

\label{t:src:fit}
\end{minipage}
\end{table*}

\begin{table*}
\scriptsize
\begin{minipage}{0.99\textwidth}
\caption{X-ray Variability of \nustar Sources}
\begin{tabular*}{\textwidth}{cc r@{}l r@{}l r@{}l r@{}l c c r@{}l r@{}l c c c}
\hline\hline
(1)                                                      & (2)      & \multicolumn{2}{c}{(3)                              } & \multicolumn{2}{c}{(4)  } & \multicolumn{2}{c}{(5) } & \multicolumn{2}{c}{(6)             } & (7)        & (8)        & \multicolumn{2}{c}{(9)                   } & \multicolumn{2}{c}{(10)             } & (11)  & (12)   & (13)  \\
No.                                                      & \chandra & \multicolumn{6}{c}{$F_X$ in 3--40 keV               } & \multicolumn{4}{c}{Flux Ratio ($r$)} & \multicolumn{4}{c}{Ran. Prob. (10\sS{-X})} & No.   & Sep.   & Var.  \\
\cmidrule(r){3-8}\cmidrule(r){9-12}\cmidrule(r){13-16}ID & ID       & \multicolumn{2}{c}{Mean                             } & \multicolumn{2}{c}{Min  } & \multicolumn{2}{c}{Max } & \multicolumn{2}{c}{Max/Min         } & $>$84.13\% & $>$99.87\% & \multicolumn{2}{c}{$f_1$ $\le$ Min       } & \multicolumn{2}{c}{$f_2/f_1$ $\ge r$} & Obs.  &        & Flags \\
(NGP)                                                    &          & \multicolumn{6}{c}{(10\sS{-6} ph s\sS{-1} cm\sS{-2})} & \multicolumn{2}{c}{                } &            &            & \multicolumn{2}{c}{$f_2$ $\ge$ Max       } & \multicolumn{2}{c}{                 } &       & (days) &       \\
\hline
2                                                        & 6090     & 1080                              & (17)    & $<$5  & .6      & 3999 & (65)    & $>$715           &    &            &            & $>$30                  & .0 & $>$8              & .0 & 3 / 5 & 364    & kr    \\
3                                                        & 4942     & 79                                & .2(4.4) & 73    & (10)    & 118  & (11)    & 1                & .6 & 1.4        & 1.0        & 6                      & .5 & 2                 & .0 & 3 / 5 & 446    &       \\
\hline
\end{tabular*}

\label{t:src:var}
Notes. (3), (4) and (5) The mean, minimum and maximum values of the
observed flux in the 3--40 keV band, respectively. (6) The
maximum-to-minimum flux ratio ($r$).  (7) and (8) An estimate for
the 1 and 3$\sigma$ equivalent lower limit of $r$, respectively,
without accounting for the multiple searches
(67 sources with multiple observations).
(9) The random chance probability ($X$ in $10^{-X}$) for flux measurements
($f_1$, $f_2$) with $f_1$ being lower than the observed minimum and
$f_2$ being higher than the observed maximum
without accounting for the multiple searches (243: the sum of the
number of observations for each source).
(10) The random chance probability 
with the ratio ($f_2$/$f_1$) being higher than the observed
the ratio ($r$) without accounting for the multiple searches (67 sources).
(11) The number of the observations used for flux calculation and the
number of the observations with the source in their FoV. The former
excludes the observations where the source falls near the chip edge.
(12) The time difference between the maximum and minimum flux measurements.
(13) Flags for short (k) and long (v) term variability. See
\S\ref{s:photometry}.
\end{minipage}
\end{table*}


\end{document}